\documentstyle[epsfig]{mn}
\begin{document}

\title
[The luminosity function of the Virgo Cluster] 
{The luminosity function of the Virgo Cluster from $M_B = -22$ to $M_B = -11$}

\author[Neil Trentham and Simon Hodgkin]  
{
Neil Trentham and Simon Hodgkin \\ 
Institute of Astronomy, Madingley Road, Cambridge, CB3 0HA. 
}
\maketitle 

\begin{abstract} 
{ 
We measure the galaxy luminosity function (LF) for the Virgo Cluster
between blue magnitudes $M_B = -22$ and $M_B = -11$ from wide-field
CCD imaging data.
The LF is only gradually rising for $-22 < M_B < -16$.  Between
$M_B = -16$ and $M_B = -14$ it rises steeply, with a logarithmic
slope $\alpha \sim -1.6$.  
Fainter than $M_B = -14$, the LF flattens again. 
This LF is shallower (although turning up at brighter 
absolute magnitudes) than the $R$-band LF recently
measured by Phillipps et al.~(1998a), who found $\alpha \sim -2.2$ fainter
than $M_R = -13$.  It is similar, however, to the LF determined from
the Virgo Cluster Catalog by Sandage et al.~(1985).   
A few faint galaxies are found which Sandage et al.~missed because their
surface-brightness threshold for detection was too high, but these
do not dominate the luminosity function at any magnitude.
Most of the faint galaxies we find are dwarf elliptical, alternatively called 
dwarf spheroidal, galaxies.  
The most important potential source of systematic
error is that we
may have rejected some high surface-brightness galaxies from the cluster
sample because we think that they are background galaxies. 
This is quite different from what has conventionally been regarded as 
the most serious source of systematic error in this kind of study:
that we are missing many {\it low} surface-brightness galaxies because
they are never visible above the sky.  

There are about 2.5 times more dwarfs per giant galaxy in Virgo 
than in the Ursa Major Cluster, a diffuse group of about
80 spiral galaxies at the same distance as Virgo, or 
the Local Group.
The Virgo and Ursa Major Cluster LFs are inconsistent with each
other at a high level of significance.  These results add weight to the
hypothesis that is developing that dwarf galaxies are
more common relative to giant galaxies in dense environments
than diffuse ones.   Both LFs are highly inconsistent with cold dark
matter theory, which has been so successful at reproducing observations
on large scales.  Possible theoretical explanations for this discrepancy,
and for the detailed shape of the Virgo Cluster LF, are investigated.  
}
\end{abstract} 


\section{Introduction} 

The Virgo Cluster is one of only two elliptical-rich clusters within 25
Mpc, the other being the Fornax Cluster.  It is near enough that even
low luminosity galaxies appear quite big on the sky
and can be identified and studied in some detail.

One arcsecond corresponds to about 0.1 kpc
in Virgo, which is comparable to the scale-lengths of low--luminosity
galaxies (Binggeli \& Cameron 1991). 
What this means is that if we see a $B>18$ galaxy in the Virgo Cluster
with a size of several arcseconds, it is much likelier to be a low luminosity
cluster member than a high luminosity background galaxy, since intrinsically
lower luminosity galaxies have lower surface-brightnesses and consequently
larger scale lengths (Binggeli 1994).   A comparison of the joint
apparent magnitude -- surface-brightness distribution between the Virgo
Cluster and blank sky fields therefore gives some indication of the 
low-luminosity galaxy
content of the Virgo Cluster. 

Sandage, Binggeli \& Tammann (1985) measured the luminosity function of
the Virgo Cluster using the Virgo Cluster Catalog (VCC; Binggeli, Sandage \&
Tammann 1985) and found a luminosity function $\phi (L)$ that was
gently rising at the faint end:
$\alpha \sim -1.35$ where  $\phi (L) \propto L^{\alpha}$. 
They detected galaxies with total $B$ magnitudes $B_{T} \sim 20$
(about $M_B =-11$), although
they estimated that their completeness limit was two magnitudes brighter
than this: at  $B_{T} > 18$, significant numbers of dwarfs have 
surface-brightnesses so low and scale-lengths so large that they were not
detectable above the night sky in the photographic images
used to compile the VCC.  Following this work, Impey, Bothun \&
Malin (1988) discovered a number of low surface-brightness
galaxies with $B_T < 20$ missing from the VCC.  They suggested that
incompleteness at the faint end of the VCC
might be very severe and that the
LF could be as steep as $\alpha = -1.7$.  In a more recent development,
Phillipps et al.~(1998a) present evidence for large numbers of low-luminosity
galaxies in the Virgo Cluster.  They measure $\alpha \sim -2.2$ fainter than
an absolute $R$ magnitude of $M_R=-13$.  
If the LF was this steep, Sandage et al.~would have missed
the vast majority of
cluster members at the faint end, even after imposing their
completeness corrections. 

All these results suggest that the LF is very likely
steeper in the Virgo Cluster
than in diffuse spiral-rich groups and clusters.  In the Local Group,
where very faint absolute magnitudes ($M_B \sim -8$) can be reached, the
faint-end slope is $\alpha = -1.1$ (van den Bergh 
1992, 2000).  In the Ursa
Major Cluster, a diffuse spiral-rich group at a similar distance to the
Virgo Cluster, less faint absolute magnitudes can be reached, but large
enough numbers of galaxies are present that a statistically robust LF can be
computed.  Here $\alpha = -1.1$ as well (Trentham, Tully \& Verheijen 2001a).
Evidence is therefore accumulating that low-luminosity galaxies are very
much more numerous per luminous galaxies in dense environments than in diffuse
ones.

Values of $\alpha \sim -2$ are of particular interest since this is
the logarithmic slope of the low-mass galaxy mass function
predicted by theory if the primordial fluctuation spectrum is a power law
with index $n=-2$, as appropriate for cold dark matter
(Press \& Schechter 1974, White \& Rees 1978,
Lee \& Shandarin 1999, Klypin et al.~1999).  If this value of $\alpha$ is
appropriate for the Virgo Cluster, this would suggest that in this
environment the efficiency of star formation in small galaxies does not
depend on the galaxy mass, assuming cold dark matter
theory  (the problem of reproducing the shallow
LF slope $\alpha \sim -1$ in the diffuse environments still remains;
Moore et al.~1999; Klypin et al.~1999). 

We now present the results of a survey of 25 square degrees 
of the Virgo Cluster (about one-tenth of the total area of the cluster)
observed through the $B$ filter using the Wide
Field Camera on the Isaac Newton Telescope on La Palma, taken as part of
the INT Wide Field Survey.  The intention is to get a reasonably complete
(at least to the surface-brightness levels defined in the 
studies mentioned above) inventory of Virgo Cluster members.  We can
then construct a luminosity function from $M_B = -22$ (the brightest galaxy
in our sample was M87 with $M_B = -21.5$) down to $M_B = -11$.  
This will permit
us to address the following questions:  
\vskip 1pt
\noindent
(i) what is the value of $\alpha$, and does this vary significantly with
absolute magnitude i.e.~over what magnitude range can we approximate the
LF by a power law?
How sensitive is the answer to our ability to recognize cluster members
based on surface-brightness: could we be missing many high surface-brightness
members because we think that they are background galaxies or many
low surface-brightness galaxies whose contrast against the sky is too low
to allow us to identify them?
\vskip 1pt
\noindent
(ii) what are the morphologies of the faintest galaxies?  Most previous
work (Sandage et al.~1985, Phillipps et al.~1998a)
suggests that they are dwarf elliptical (alternatively called
dwarf spheroidal) galaxies.
This would imply that low surface-brightness irregular star-forming
galaxies do not contribute significantly to the LF at the faint end
as they do in the Ursa Major Cluster (Trentham et al.~2001a);
\vskip 1pt
\noindent
(iii) do the results depend on the colour of the galaxies and the filter
used?  We, like Sandage et al., are using a $B$ filter.  Phillipps 
et al.~used a red $R$ filter. 
How much of the very substantial excess of galaxies they found could be
due do this, given that dSph/dE galaxies tend to be red (Caldwell 1983)?
\vskip 1pt
\noindent
(iv) are there substantial numbers of very low surface brightness (VLSB)
galaxies, as seen in the Fornax cluster by Kambas et al.~(2000)?
In particular, as we approach the limiting surface brightness
at which we can detect objects in our data, do we find more and more
VLSB galaxies, and what is the their contribution
to the total LF?   We will need to quantify this
in order to address the questions posed in (i) above; 
\vskip 1pt
\noindent
(v) how does the galaxy luminosity function depend on environment?  We will
compare our Virgo Cluster luminosity function to the $B$-band luminosity
functions of the Local Group, the Ursa Major Cluster, and the rich
Coma Cluster at a distance of 90 Mpc.  We will then have measurements of the
LF in four very different environments.
In the case of the Local Group, the 
LF has large uncertainties due to Poisson statistics.  In the Coma 
Cluster the LF has large uncertainties due to the need to do a background
subtraction.  It is well-known (Dressler 1980) that the morphologies of
galaxies depend on environment, specifically on the galaxy density.  Our
results will indicate whether or not the LF does too, over a large
magnitude range.

\section{Observations} 

\begin{figure}
\begin{center}
\vskip-4mm
\psfig{file=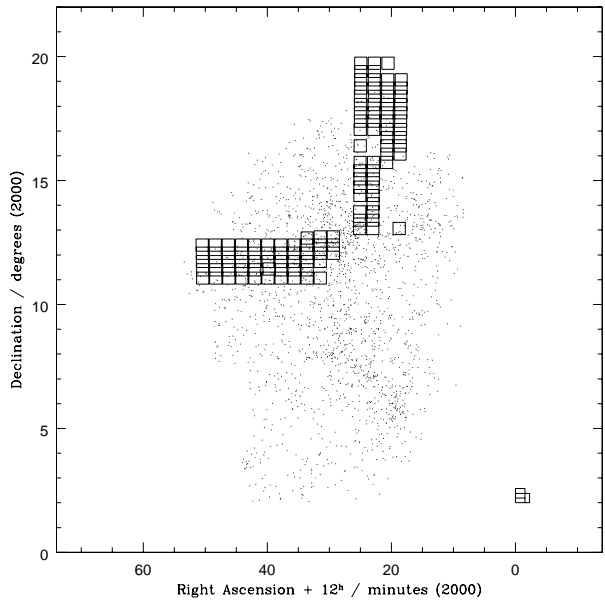, width=8.65cm}
\end{center}
\vskip-3mm
\caption{
The region of the Virgo Cluster studied.  The open
boxes represent the 113 fields observed; the box size corresponds approximately
to the field of view of the INT Wide Field Camera.  The dots represent the
locations of the 2096 Virgo Cluster Catalog galaxies (Binggeli et al.~1995). 
The exact shape of the INT Wide Field Camera, indicating the positions of
the 4 CCDs, is shown in the bottom right hand corner.}
\end{figure}

The data used here were taken on various observing runs during Spring 1999
and Spring 2000 as part of the INT Wide Field Survey
(WFS; http://www.ast.cam.ac.uk/$^{\sim}$wfcsur; McMahon et al.~2001).
This is a digital survey covering about
100 deg$^{2}$ of sky carried out using the Wide Field Camera
(a mosaic of four 4K $\times$ 2K EEV CCDs, pixel scale 0.33 arcsec
pix$^{-1}$, field of view 0.29 deg$^{-2}$;
http://www.ast.cam.ac.uk/$^{\sim}$wfcsur/ccd.html) on the 2.5 m Isaac
Newton Telescope on La Palma.

A total of 113 fields in the Virgo Cluster
were selected (see Fig.~1), comprising 24.9 deg$^2$ taking
into account overlaps between the images,
which were typically 1 chip or 1/4 of
the field of view.
All fields were imaged through a $B$ (effective
wavelength $\lambda_0 = 0.44$ $\mu$m) filter for 750 s
under photometric conditions.
A total of 31 deg$^2$
from the WFS imaging survey of the North Galactic Cap (NGC),
taken in photometric conditions with a $B$ filter, were also studied, these
being used as offset fields to
estimate the background contamination in the Virgo
sample.  The exposure times in $B$ (750 s) were the same for the Virgo and NGC
data.

The data were pre-processed and reduced via the WFS pipeline
(http://www.ast.cam.ac.uk/$^{\sim}$wfcsur/ pipeline.html;
Irwin \& Lewis 2001) and photometric
calibration obtained from observations of several (5 -- 10
per night) standard
stars.  The photometric zero points were always accurate to 2 \%; uncertainties
in the zero point are
not a major source of uncertainty in the galaxy magnitudes
that we derive.
The median seeing was 1.95 arcseconds
for the Virgo data and 1.26 arcseconds for the
NGC data (since we will be searching for galaxies with large sizes and low
surface-brightnesses,
the relatively poor seeing of the Virgo data is not a major
concern).  For the Virgo data, the median 3$\sigma$
point source limiting magnitude
was $B=25$.

Of the 113 fields we studied,  9
of the fields were also imaged through a $Z$ ($\lambda_0 = 0.90$ $\mu$m) filter
for either 600s or 1200 s under photometric conditions.  This data was used
to compute $B-Z$ colours for a small number of galaxies in the $B$ sample.
The median seeing in the $Z$ dataset was 1.53 arcseconds FWHM.

\section{Galaxy identification and selection}

\begin{figure*}
\begin{minipage}{170mm}
{\vskip-3.5cm}
\begin{center}
\epsfig{file=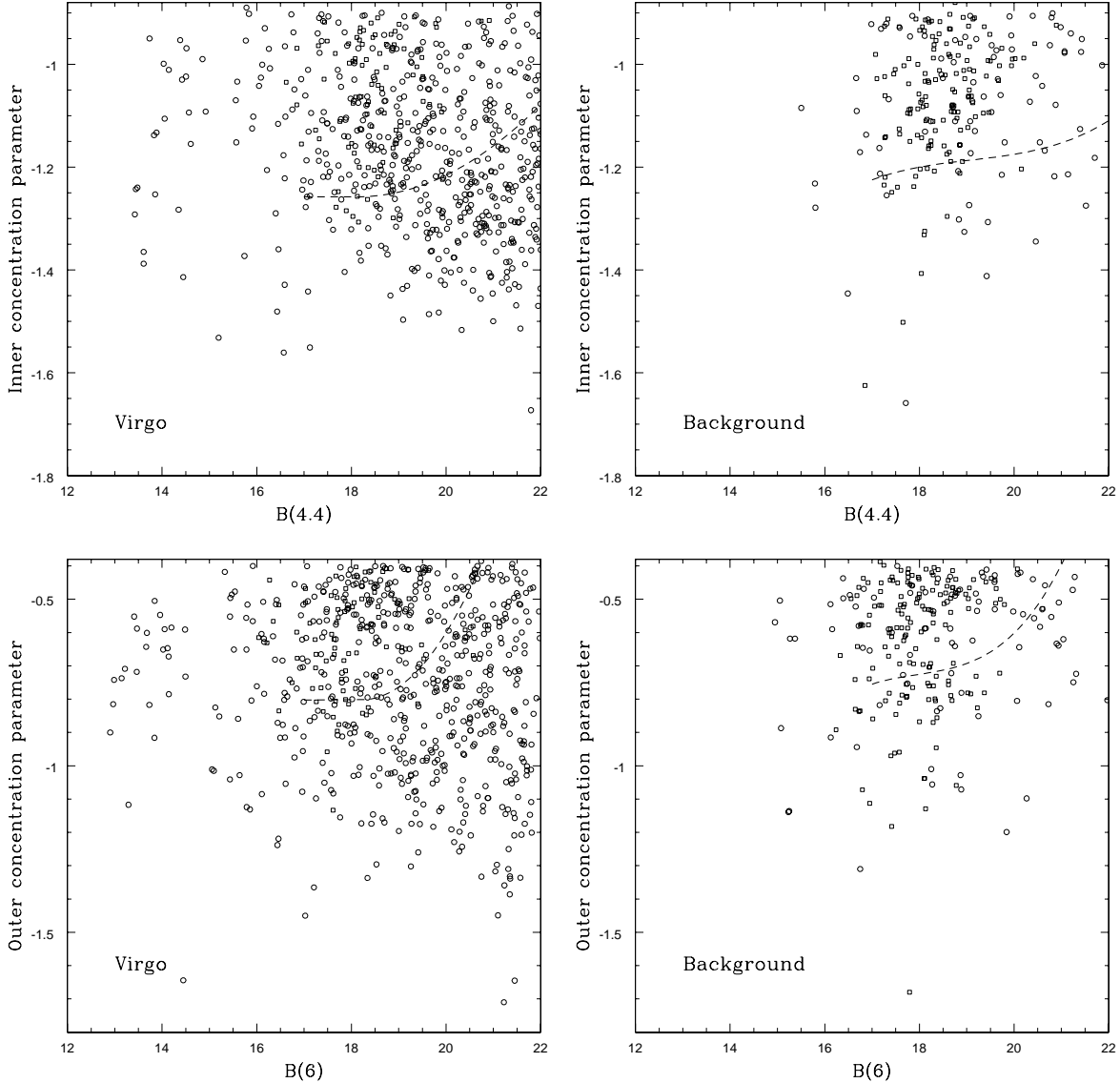, width=18.65cm}
\end{center}
{\vskip-5.7cm}
\caption{Inner and outer concentration parameters as a function of
aperture magnitude 
for all objects we detected in both the Virgo and NGC
(background) data that did not have close
companions projected within 12 arcseconds.  Objects indicated with a
square (as opposed to a circle) showed either clear spiral structure or
were highly flattened disks, usually with central bulges; these
are almost certainly background objects if seen in the Virgo fields. 
The dashed lines represent 
simulated typical dwarfs from Fig.~1 of Binggeli (1994; surface-brightnesses
$\mu_B$ in mag arcsec$^{-2}$ and absolute blue magnitudes 
$M_B$ are related by the
approximation $\mu_B \approx 33 + 0.59 M_B$) 
convolved with the seeing and ``observed'' in the presence of noise
appropriate to the data.  Exponential profiles and axial ratios of 1 were
assumed for the simulated dwarfs.}
\end{minipage}
\end{figure*}

Initially an image detection algorithim developed by Irwin (1985,
1996) was run on the images.  A source was defined to be 5 contiguous
pixels with a flux signal-to-noise ratio of 1.5 above the sky.
Generated parameters include the position, intensity and shape
of each detected object.
We used only the position information.
Aperture magnitudes were
subsequently calculated for all the sources using the IRAF
{\it APPHOT} task with a local sky value measured for each object (typically
the mode within an annulus of inner and outer radii 0.5 and 1.0 arcminutes
centered on the object).
 
In our dataset we had about $10^6$ galaxies (about 2500/chip) with
$B$ magnitudes within a 6 arcsecond aperture brighter than $B(6) = 23$
(here, as throughout the paper we use the notation
$B(r)$ to mean the magnitude within an aperture of radius $r$ arcseconds).
In this section we describe how we identified which ones are likely to be
Virgo Cluster members and assess how much confidence we should have in these
identifications.

Many of the brightest galaxies (typically those with $B(6) < 18$)
had spectroscopic velocity measurements available.  For such objects
we can identify with 100 \% confidence which are members and which are not.
We follow Binggeli et al.~(1985) in assigning cluster membership to all
galaxies with heliocentric velocities between $-$700 km s$^{-1}$ and
2700 km s$^{-1}$.

For the other galaxies, the situation is more complex.  As outlined in
Section 1, we expect dwarf galaxies in the cluster to have lower
surface brightnesses (and consequently more extended light profiles) than
background galaxies of the same apparent magnitudes.  Morphologically
we expect them to be either smooth (dwarf elliptical/spheroidal galaxies) 
or lumpy (dwarf irregulars) but {\it not} to have appreciable spiral
structure or a distinctive bulge+disk morphology, both of which
are characteristic of luminous giant galaxies (Binney \& Merrifield 1997)
which would here be background galaxies.  

We used both surface-brightness and morphology
to assess membership as follows.
Firstly we compared the light concentrations of galaxies in the
Virgo and background samples.  For each galaxy we computed (following
Trentham et al.~2001a) an inner concentration parameter (ICP) and an
outer concentration parameter (OCP), defined as
$${\rm ICP} = B(4.4) - B(2.2)$$ 
$${\rm OCP} = B(12) - B(6).$$ 
The results are presented in Figure 2. 

There is a clear excess of objects with $B(4.4) > 18$ and ICP
$\sim - 1.3$ in the Virgo fields that are not seen in the background
fields.  These are presumably Virgo cluster members.  They have the
correct sizes for dwarf galaxies seen at the distance of the Virgo Cluster 
since they are close to the dashed lines in the upper left panel of Figure 2.
The scatter around the dashed lines is huge, however, particularly in
the lower panels.  The dependence on surface
brightness of the location of galaxies in the panels in Figure 2 is
complex.  In general lower surface brightness galaxies have more extended
light profiles yet 
it is not {\it always} true that lower surface-brightness galaxies 
always have more negative concentration parameters.
In the lower panels very low surface
brightness galaxies have very small differences between
their 6-arcsecond and 12-arcsecond aperture fluxes since so much of
the galaxy falls below the sky; if the galaxy is faint (as for galaxies at the
right end of the dashed lines), the only difference between $B(6)$ and
$B(12)$ is due to sky noise.  This could lead to an OCP value
that could be as high as $\infty$!  Hence there is huge scatter 
at the faint end
in the lower two panels.  What all this means is that we must consider
the $B(6)$ and concentration parameter values in conjunction with each
other when assessing
membership i.e.~in effect we must consider the entire light profile.     

We therefore define a quantity
\begin{equation}
P = { {
v({\rm OCP},B(6)) - b({\rm OCP},B(6))/1.25
}\over{v({\rm OCP},B(6)) } } 
\end{equation}
where $v$(OCP$^{\prime}$,$B$(6)$^{\prime}$) is the number of 
objects in the Virgo sample with
6-arcsecond aperture magnitudes within 0.5 mag of $B(6)^{\prime}$ and an
outer concentration parameter within 0.05 mag of OCP$^{\prime}$ 
{\it and an ICP value lower than $-$0.9 mag} and
$b$(OCP$^{\prime}$,B(6)$^{\prime}$) is the equivalent number for the
background fields.  
The factor of 1.25 comes from the relative area samples by our Virgo
and background surveys. 
Defined this way $P$ is then an estimate of the probability that a particular
galaxy is a cluster member or not.  

For cluster members with velocity measurements, $P \sim 1$ so we can have
some confidence in using $P$ values to assess membership.  But some
grand-design spirals which are clearly background (these have very
negative OCPs due to considerable amounts of star formation far from the
galaxy centres) also have $P \sim 1$.  
Hence using $P$ values as a direct measure of membership is
not completely safe.  Other (related) reasons why we might be suspicious are:
\vskip 1pt \noindent
(1) The scatter in Fig.~2 around the dashed lines is large
(see the previous discussion), so on an
object-by-object basis, the uncertainty in $P$ is large.
\vskip 1pt \noindent
(2) Our background sample (even though it comprises about 30
deg$^2$ of data) is not sufficiently large to compensate for shot noise
and Poisson errors in $b$(OCP$^{\prime}$,B(6)$^{\prime}$) in regions
where the OCP values are large.  For example, in the lower right panel of 
Figure 2, there are only three galaxies with an OCP $<-1.1$.  These are
all low surface-brightness field galaxies.    
\vskip 1pt \noindent
(3) The field-to-field variance of these low surface-brightness field galaxies
is determined by large-scale structure at very low redshift and is consequently
likely to be large.  This makes the problem highlighted in (2) above more
severe.  For example, suppose for some combination of $B(6)$ and OCP,
we have 10 Virgo field galaxies with ICP $ <-0.9$ mag and 3 background
field galaxies.  We would compute $P = (10 - 3/1.25)/10 = 0.81$.  We
would therefore think that most of these galaxies are cluster members.
But suppose large-scale structure at low redshift caused the field-to-field
variance of these low surface-brightness galaxies to be a factor of four. 
Then the appropriate number of background galaxies with the relevant
combination of parameters in the Virgo fields could now be
twelve, not three.     
So $P$ should have been $(10 - 12/1.25)/10 = 0.04$, which is negligible.
With this value of
$P$, we would now think that most of these galaxies were {\it not} cluster
members.  In summary, until we have a characterization of the
field-to-field variance of field galaxies {\it as a function of
surface-brightness}, values of $P$ must be treated with caution. 

So instead of relying solely on $P$ values, we further inspected each galaxy
individually and made a judgment about the possibility of membership.
We discarded galaxies showing obvious spiral structure and flat disk
galaxies with a central bulge, even if their $P$ values were high,
since these are almost certainly background giant galaxies (about
15\% of the galaxies that we would have  
identified as possible members in the absence of any
morphological information beyond averaged light profiles 
were rejected on these
grounds).  For each
galaxy that did not have a velocity measurement that we thought might
be a member we rated it (as was
done in Trentham et al.~2001a) ``1'' or ``2'' depending on whether we
thought it was a member with a high or moderate degree of confidence.
The rating scheme is summarized in Table 1.

In making this analysis, we relied for the main part on $P$ values, but also
on morphology, particularly in regions of parameter space which were
poorly sampled in the background fields (see Point (2) above).  Most
objects rated ``2'' were galaxies of moderate surface-brightness where there
was some overlap in parameter space between the background and Virgo; such
galaxies were not rated ``1'' due to point (3) above.   

\begin{table*}
\caption{Rating scheme for Virgo galaxies}
{\vskip 0.75mm}
{$$\vbox{
\halign {\hfil #\hfil && \quad #\hfil \hfil \cr
\noalign{\hrule \medskip}
Rating & Meaning & Comments &\cr
\noalign{\smallskip \hrule \smallskip}
\cr
0 & Confirmed member & Velocity measurements exist &\cr 
  &                  & Heliocentric velocity between $-$700 km s$^{-1}$ 
                       and 2700 km s$^{-1}$ &\cr
\cr
1 & Probable member & Velocity measurements do not exist; &\cr 
  &                 & High $P$; &\cr
  &                 & Very low surface brightness  &\cr 
\cr
2 & Possible member & Velocity measurements do not exist; &\cr 
  &                 & Moderate $P$; &\cr
  &                 & Low surface brightness, but not lower than
                        the lowest 
                      surface-brightness background galaxies  &\cr 
\noalign{\smallskip \hrule}
\noalign{\smallskip}\cr}}$$}
\end{table*}

There were a few additional complications resulting from this approach
that required attention.  
\vskip 3pt \noindent
1) The Virgo images were taken under a variety of seeing conditions.  In
the worst conditions the seeing was 3 arcseconds.  
Galaxies in these images are slightly bigger and have 
less concentrated light profiles than galaxies in Virgo images taken under more
typical conditions or in the background fields.  
We simulated galaxies and found that 
differences in $P$ due to seeing variations of this magnitude
were negligible relative to other uncertainties; this is not 
surprising given that we are selecting low surface brightness galaxies
that are typically much bigger than 3 arcseconds;  
\vskip 3pt \noindent
2) Occasionally galaxies of interest fell on cosmic rays or chip
defects.  Since all candidates on the images are being studied
by eye (to assign ratings), it was easy to identify when this happened.
In these cases, the cosmic ray or defect was removed by interpolation using
the IRAF {\it imedit} task.  All subsequent analysis was performed on
the repaired images;  
\vskip 3pt \noindent
3) Some objects appeared in more than one exposure (the fields in Fig.~1
overlap).  Occasionally these had very different concentration parameters
from each other: one or the other might have fallen near a chip edge or
defect.  In this event we only used the measured parameters for the undamaged
image.  Otherwise data from both exposures were combined and used;
\vskip 3pt \noindent
4) Often, on the initial detection pass, very low surface-brightness
objects were detected as several separate objects, each centered on a local
noise peak.  These objects were identified upon inspection and all
objects except the single object centered on the galaxy center removed
from the catalogue.  Subsequent photometry performed on this single object;
\vskip 3pt \noindent
5) Similarly, on occasion where galaxies (typically dwarf irregular)
had off-center hot-spots, the initial detection algorithm identified
objects centered on these hotspots.  In these cases, we recentered the
galaxy and performed subsequent photometry on this recentered object;
\vskip 3pt \noindent
6) Many objects have close companions that were identified upon inspection.
The $B(6)$ concentration parameters for such objects have little meaning.  
Total magnitudes were determined individually as described in
Section 4;
\vskip 3pt \noindent
7) A few objects, like IC 3483 (heliocentric velocity 108 km s$^{-1}$),
fell in the gaps between the CCDs and are missing from our catalogue,
even though they are in the area outlined in Figure 1.  This may also
have happened for some lower luminosity galaxies.  The survey area we
quoted in the previous section includes a (small) correction for area
lost in this way;  
\vskip 3pt \noindent
8) Two very high surface-brightness galaxies that look like background galaxies
but have velocities which place them in the Virgo
Cluster were included (a search for all objects with known velocities
was performed using the NASA/IPAC Extragalactic Database, hereafter NED).
The two objects were VCC 1313 (a blue compact dwarf) and VCC 1627 (a
compact low-luminosity elliptical galaxy).  Both lack a detectable
low surface-brightness halo at large galactocentric radius, normally
a defining characteristic of low-luminosity galaxies.  Such galaxies are
not common for $B < 17$ ($M_B < -14$),
but were they to be very common at $B > 17$, they
would be missing from our
sample and the LF that we determine would be incomplete at the faint end.
We do not, however, regard this possibility as likely and return to this
issue in Section 7;  
\vskip 3pt \noindent
9) Low surface-brightness features likely to be associated with luminous
galaxies (whether cluster members or background) were judged not to be
independent galaxies and removed from the catalogue.  For example, there
is a large stream of material flowing from the background galaxy pair
IC 3481/A (heliocentric velocity $\sim 7000$ km s$^{-1}$) that we
excluded because it is likely to be associated with these galaxies and
not the Virgo Cluster.  
Similarly, M100 in the cluster is surrounded by considerable amounts of
low surface-brightness material that we assume is associated with that
galaxy and is not made up of individual cluster dwarfs; 
\vskip 3pt \noindent
10)  For low surface-brightness objects that fell near the edge of a CCD,
it was sometimes not clear if the centre of the object fell on the CCD 
or not and it is possible that we included objects that we should have
rejected under point 7) above.   However this happened rarely enough that
it is not a significant source of error.   

\section{Galaxy sample}

Our sample consists of 449 galaxies of which we rated
102 as ``0'', 220 as ``1'', and 127 as ``2''.
Images of typical galaxies in our sample are presented in Fig.~3. 

\begin{figure*}
\begin{minipage}{170mm}
{\vskip-3.5cm} \begin{center} \epsfig{file=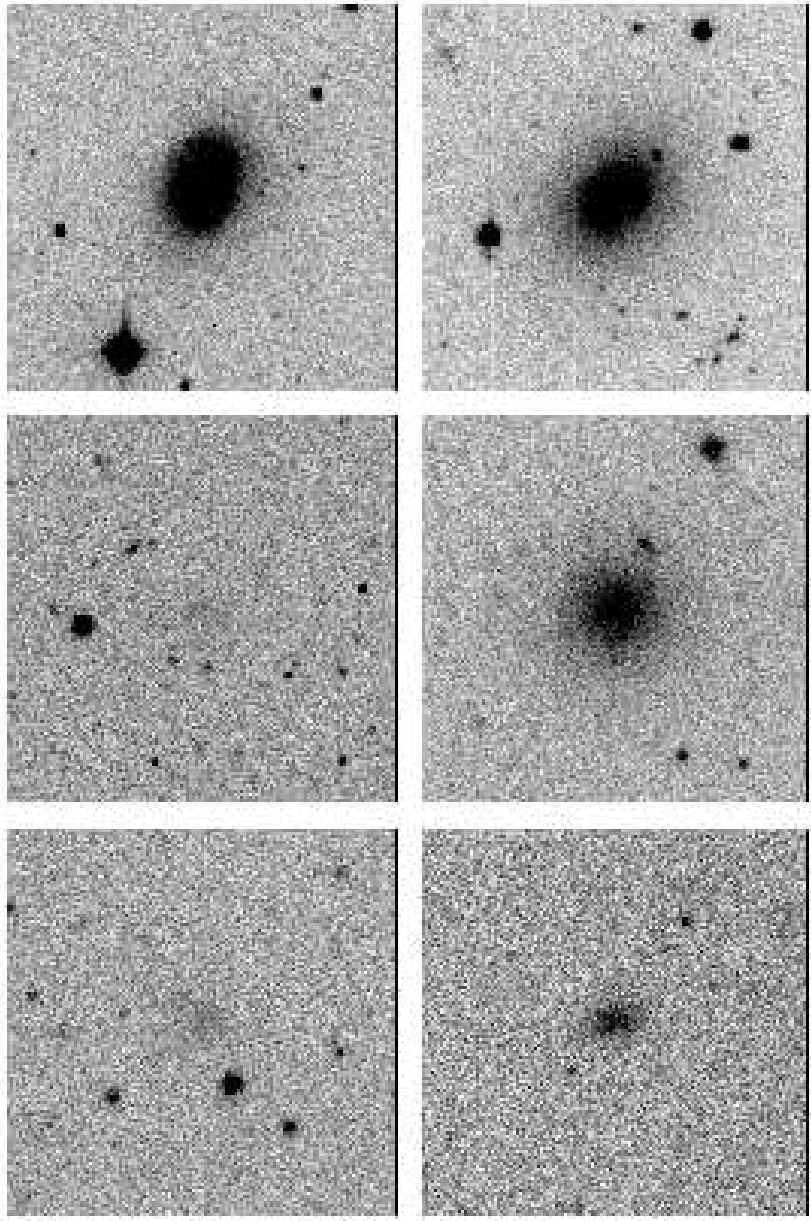, width=18.65cm}
\end{center}
{\vskip-4.9cm}
\caption{Sample galaxies from the catalogue, having $B(6)$ 
values of 17 through 22 in 1 magnitude units, clockwise from top
left.  All images are square, 132 arcseconds on a side.} 
\end{minipage}
\end{figure*}

The galaxy sample is presented in Table 2.  
There we list:
\vskip 1pt \noindent
(1) the galaxy identification number, ranked by $M_B$. 
A single
star by the name means that we identified a  
companion close enough that it would affect photometry
measurements at radii $6^{\prime \prime} < r <
12^{\prime \prime}$ from the galaxy centre but not
at $r < 6^{\prime \prime}$.   
Measurements of the OCP and $P$ for these galaxies
are consequently not an indicator of the galaxy
properties.  A double 
star by the name means that we identified a
companion close enough that it would affect photometry
measurements at $r <
6^{\prime \prime}$.
Measurements of the $B(6)$, the ICP,
OCP and $P$ for these galaxies
are consequently not an indicator of the galaxy
properties for these objects;
\vskip 1pt \noindent
(2) the galaxy name as appearing in the
NASA/IPAC Extragalactic Database (NED).  The VCC numbers
refer to the Virgo Cluster Catalog
(Binggeli et al.~(1985) and the IBM88 designations
refer to objects studied by Impey et al.~(1988); 
\vskip 1pt \noindent
(3) the galaxy type, taken either from the Revised
Third Catalog (RC3; de Vaucouleurs et al.~1991) of
galaxies, the VCC, or our own inspection of the 
images.  The designations are E = elliptical galaxy,
S0 = lenticular galaxy,
S = spiral galaxy, with following letters indicating
the Hubble type, BCD = blue compact dwarf,
VLSB = very low surface brightness galaxy (these were labeled as such
based on a visual inspection and tended correspond to galaxies having
average surface brightnesses within an aperture of radius 12 arcseconds
fainter than 26.5 $B$ mag arcsec$^{-2}$; at the very faint end of the
sample, this kind of classification becomes tenuous and the distinction
between VLSB galaxies and other dwarfs is purely subjective), dE =
dwarf elliptical/dwarf spheroidal galaxy, dS0 =
dwarf lenticular, dI = dwarf irregular.  The notation
dE/I means that we were unable to identify from the
images what the type of dwarf galaxy was;
\vskip 1pt \noindent
(4) right ascension;
\vskip 1pt \noindent
(5) declination; 
\vskip 1pt \noindent
(6) heliocentric radial velocity.  See NED for
the original references;
\vskip 1pt \noindent
(7) $B(6)$, the apparent blue magnitude measured
within a 6$^{\prime \prime}$ radius;
\vskip 1pt \noindent
(8) the ICP, or inner concentration parameter,
defined as in Section 3.  This is not meaningful for objects with  
companions;
\vskip 1pt \noindent
(9) the OCP, or inner concentration parameter,
defined as in Section 3;
\vskip 1pt \noindent
(10) the total apparent blue magnitude $B_T$.  For
the very brightest galaxies ($B_T < 15$) we adopt the RC3
values when available, since these galaxies normally extended
beyond the edges of a single CCD chip in our data.
For the majority of galaxies we computed $B_T$ from the 
magnitudes within a 12 arcsecond aperture, corrected
for light at large radius which has fallen below the
sky using the prescription of Tully et al.~(1996).  
This method requires us to assume an azimuthally-averaged exponential
profile $B(r) = -2.5 \log_{10} \int_0^r \, I_0 \, 
\exp (-r/h) \, 2 \pi r \, {\rm d}r$,
to derive the scale-length
$h$ and central intensity $I_0$ from a fit at radii below 12 arcseconds, and
then to compute $B_T = B({\infty})$.  We estimate errors of about 0.2
magnitudes in this analysis, derived from comparing $B_T$ values derived
from objects appearing in more than one image. 
More serious systematic errors might result if the light at large radii
that fell below the sky is {\it not} exponential or follows a different
exponential law from what we derive by fitting at smaller radii; however there
is no evidence for this phenomenon happening in the typical dwarf galaxies 
observed by Binggeli \& Cameron (1991). 
The only galaxy where we did not use an
exponential profile for this extrapolation was the elliptical galaxy NGC 4486A, 
where we used an $r^{1/4}$-law
extrapolation (de Vaucouleurs 1948).  
For objects labeled ``**'' the companions prevent a fit from
giving meaningful $h$ and $I_0$ values (this effect was not
significant for objects labeled ``*''), so the $B_T$ values were derived
individually by identifying a symmetry axis and computing the flux  
from the part of the galaxy on the other side of this axis from
the companion and multiplying
this flux by two.  Additionally, VCC 2062 was treated this way, since the
light from this irregular galaxy peaks so far away from its centre. 

Since we will ultimately consider the LF for the entire
magnitude range $-22 < M_B < -11$, it is important to consider
how the RC3 $B_T$ magnitudes used at very bright magnitudes  
compare with the $B_T$ magnitudes we derive from the
exponential fitting method at fainter magnitudes.
We made the transition from using the different types
of measurements at $B_T = 15$.  For objects within 0.5 magnitude
of this transition magnitude, the difference $\Delta$
between the $B_T$ values obtained from the RC3 and those we derive
is $\Delta = 0.16 \pm 0.10$ mag, which is smaller than our
expected uncertainty in the $B_T$ values we derive (see above)
and much smaller than the binwidth (1 mag) we use to compute
the LF.  The RC3 magnitudes will be more accurate for brighter galaxies,
but even at the faintest limits that we use them, they 
therefore appear to
be accurate enough for our purposes. 

\begin{figure}
\begin{center}
\vskip-4mm
\psfig{file=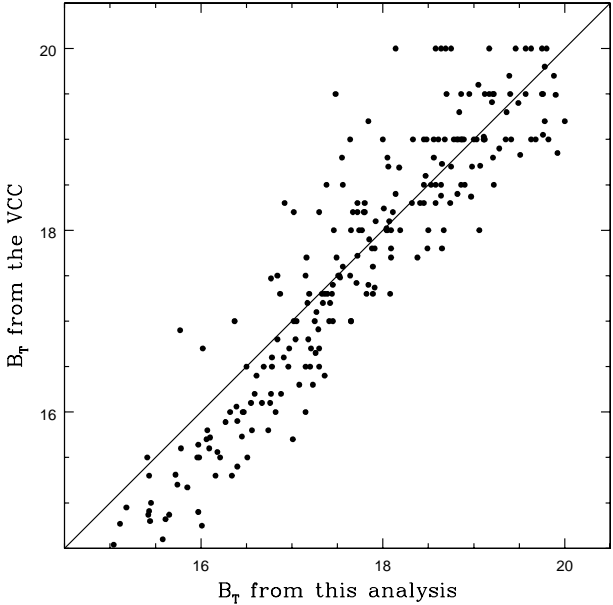, width=8.65cm}
\end{center}
\vskip-3mm
\caption{
Comparison between the $B_T$ magnitudes derived in the present
analysis and those from the VCC (Binggeli et al.~1995).
}
\end{figure}

It is also instructive to compare our derived magnitudes
with the magnitudes in the VCC for all dwarfs common to both
catalogues (see Fig.~4).  
For $M_B = -16$ to
$M_B = -11$ we have 318 galaxies, 232 (73 \%) of which are VCC members.
Some are not (like UGC 7346) since they are not in the
VCC fields -- in this study we observed fields beyond the northern
extent of the VCC (see Fig.~1) survey area. 
Some are not VCC members (like the IBM88 objects) because their
surface brightnesses were too low to be included in that sample.
From Fig,~3, we see that the VCC magnitudes brighter
than $B_T = 18$ tend to be
systematically brighter than ours by about half a
magnitude (a similar, albeit smaller,
effect was noted by Binggeli et al.~1995
who compared the VCC magnitudes with the $B_T$ values derived by
Ichikawa et al.~1986).  Fainter than $B_T = 17$ there is no
systematic offset between the VCC magnitudes and ours
(the mean difference is 
$B_T$ (present) $-$ $B_T$ (VCC) =
$-$0.01 $\pm$ 0.04 mag).  
\vskip 1pt \noindent
(11) the nominal probability $P$ of being a cluster
member, as determined in Section 3 from a comparison
with the background fields;
\vskip 1pt \noindent
(12) Galaxy rating according to the scheme described in Table 1; 
\vskip 1pt \noindent
(13) absolute magnitude $M_B = B_T - 31.15 - A_B$,
where the Galactic Extinction 
$A_B$ values (typically 0.1 -- 0.2 magnitudes) are adopted from NED
and based on the measurements of Schlegel
et al.~(1998).
We adopt a distance modulus of 31.15 for the Virgo Cluster (Tonry
et al.~2001). 

\begin{table*} \caption{The Sample} {\vskip 0.55mm} {$$\vbox{ \halign {\hfil #\hfil && \quad \hfil #\hfil \cr
\noalign{\hrule \medskip}
(1)  & (2) &  (3) & (4) & (5) & (6) &
(7)  & (8) & (9) & (10) &
(11) & (12) & (13) &\cr
ID  & Name &  Type & $\alpha$ (J2000) & $\delta$ (J2000) & $V_h$/km s$^{-1}$ &
$B(6)$  & ICP & OCP & $B_T$ &
$P$ & Class & $M_B$ &\cr
\noalign{\smallskip \hrule \smallskip}
   1 & M87 (NGC 4486) & E &  12 30 49.46 & 12 23 28.8 & 1307   & 13.29 & $-1.25$ & $-1.12$ &  9.59 & 1.00 & 0 & $-21.64$ &\cr
   2 & M60 (NGC 4649) & E &  12 43 38.02 & 11 33 11.2 & 1117   & 14.45 & $-1.51$ & $-1.56$ &  9.81 & 1.00 & 0 & $-21.45$ &\cr
   3 & M86 (NGC 4406) & E &  12 26 11.64 & 12 56 56.6 & $-$244 & 15.64 & $-1.53$ & $-1.64$ &  9.83 & 1.00 & 0 & $-21.45$  &\cr
   4 & M85 (NGC 4382) & S0   & 12 25 23.43 & 18 11 26.7 &  729 & 12.99 & $-1.24$ & $-0.74$ & 10.00 & 1.00 & 0 & $-21.28$  &\cr
   5 & M84 (NGC 4374) & E    & 12 25 03.76 & 12 53 13.4 & 1060 & 12.97 & $-1.24$ & $-0.82$ & 10.09 & 1.00 & 0 & $-21.23$  &\cr
   6 & M100 (NGC 4321)& Sbc  & 12 22 54.74 & 15 49 20.5 & 1571 & 13.84 & $-1.41$ & $-0.92$ & 10.05 & 1.00 & 0 & $-21.21$  &\cr
   7 & M58 (NGC 4579) & Sb   & 12 37 43.56 & 11 49 05.6 & 1519 & 13.47 & $-1.13$ & $-0.72$ & 10.48 & 1.00 & 0 & $-20.85$  &\cr
   8 & M59 (NGC 4621) & E    & 12 42 02.25 & 11 38 48.8 &  410 & 13.21 & $-1.39$ & $-0.71$ & 10.57 & 1.00 & 0 & $-20.72$  &\cr
   9 & M89 (NGC 4552) & E    & 12 35 39.83 & 12 33 23.4 &  340 & 13.14 & $-1.37$ & $-0.74$ & 10.73 & 1.00 & 0 & $-20.60$  &\cr
  10 & NGC 4293 & S0/a & 12 21 12.84 & 18 22 58.6 &  893 & 15.09 & $-1.03$ & $-1.02$ & 11.26 & 1.00 & 0 & $-20.06$  &\cr
  11 & NGC 4568 & Sbc  & 12 36 34.30 & 11 14 20.8 & 2255 & 15.43 & $-1.13$ & $-1.04$ & 11.68 & 1.00 & 0 & $-19.61$  &\cr
  12 & NGC 4394 & Sb   & 12 25 55.67 & 18 12 50.4 &  922 & 14.20 & $-1.09$ & $-0.59$ & 11.73 & 1.00 & 0 & $-19.55$  &\cr
  13 & NGC 4567 & Sbc  & 12 36 32.71 & 11 15 28.4 & 2274 & 17.21 & $-1.26$ & $-1.37$ & 11.79 & 1.00 & 0 & $-19.40$  &\cr
  14 & NGC 4647 & Sc   & 12 43 32.31 & 11 34 54.7 & 1422 & 15.06 & $-1.15$ & $-1.01$ & 11.94 & 1.00 & 0 & $-19.32$  &\cr
  15 & NGC 4503 & S0   & 12 32 06.26 & 11 10 35.5 & 1342 & 14.02 & $-0.95$ & $-0.65$ & 12.05 & 1.00 & 0 & $-19.32$  &\cr
  16 & NGC 4694 & S0   & 12 48 15.14 & 10 59 00.1 & 1175 & 14.14 & $-0.97$ & $-0.67$ & 12.06 & 1.00 & 0 & $-19.26$  &\cr
  17 & NGC 4564 & E    & 12 36 27.02 & 11 26 21.7 & 1142 & 13.68 & $-1.00$ & $-0.60$ & 12.05 & 1.00 & 0 & $-19.25$  &\cr
  18 & NGC 4567 & Sab  & 12 36 33.29 & 11 15 45.0 & 2274 & 16.44 & $-1.23$ & $-1.24$ & 12.06 & 1.00 & 0 & $-19.23$  &\cr
  19 & NGC 4660 & E    & 12 44 32.01 & 11 11 25.9 & 1083 & 13.47 & $-1.14$ & $-0.59$ & 12.16 & 1.00 & 0 & $-19.13$  &\cr
  20 & NGC 4638 & S0   & 12 42 47.38 & 11 26 32.4 & 1164 & 13.66 & $-1.11$ & $-0.64$ & 12.13 & 1.00 & 0 & $-19.13$  &\cr
  21 & NGC 4478 & E    & 12 30 17.46 & 12 19 43.2 & 1349 & 13.73 & $-1.01$ & $-0.82$ & 12.36 & 1.00 & 0 & $-18.90$  &\cr
  22 & NGC 4486A & E   & 12 30 57.76 & 12 16 14.5 &  450 & 13.41 & $-0.95$ & $-0.55$ & 12.44 & 1.00 & 0 & $-18.81$  &\cr
  23 & NGC 4550 & S0   & 12 35 30.62 & 12 13 15.0 &  381 & 14.14 & $-1.16$ & $-0.79$ & 12.56 & 1.00 & 0 & $-18.76$  &\cr
  24 & NGC 4312 & Sab  & 12 22 31.39 & 15 32 15.9 &  153 & 15.86 & $-1.29$ & $-1.13$ & 12.53 & 1.00 & 0 & $-18.74$  &\cr
  25 & NGC 4402 & Sb   & 12 26 07.01 & 13 06 48.4 &  232 & 17.02 & $-1.32$ & $-1.45$ & 12.55 & 1.00 & 0 & $-18.73$  &\cr
  26 & NGC 4379 & S0   & 12 25 14.78 & 15 36 26.9 & 1069 & 14.11 & $-1.02$ & $-0.65$ & 12.63 & 1.00 & 0 & $-18.62$  &\cr
  27 & NGC 4606 & Sa   & 12 40 57.48 & 11 54 42.1 & 1664 & 15.21 & $-1.37$ & $-0.85$ & 12.67 & 1.00 & 0 & $-18.62$  &\cr
  28 & NGC 4383 & S0   & 12 25 25.53 & 16 28 12.8 & 1710 & 13.84 & $-0.83$ & $-0.51$ & 12.67 & 1.00 & 0 & $-18.58$  &\cr
  29 & NGC 4377 & S0   & 12 25 12.46 & 14 45 42.0 & 1371 & 13.96 & $-1.28$ & $-0.55$ & 12.76 & 1.00 & 0 & $-18.55$  &\cr
  30 & NGC 4733 & S0/a & 12 51 06.80 & 10 54 43.6 &  908 & 15.12 & $-1.07$ & $-0.83$ & 12.70 & 1.00 & 0 & $-18.54$  &\cr
  31 & NGC 4528 & S0   & 12 34 06.10 & 11 19 16.9 & 1374 & 14.05 & $-1.03$ & $-0.59$ & 12.97 & 1.00 & 0 & $-18.38$  &\cr
  32 & NGC 4551 & S    & 12 35 37.98 & 12 15 50.7 & 1172 & 14.49 & $-1.09$ & $-0.73$ & 12.97 & 1.00 & 0 & $-18.35$  &\cr
  33 & NGC 4476 & S0   & 12 29 59.14 & 12 20 55.6 & 1978 & 14.48 & $-0.99$ & $-0.59$ & 13.01 & 1.00 & 0 & $-18.26$  &\cr
  34 & NGC 4396 & Sd   & 12 25 59.20 & 15 40 14.9 & $-$128 & 16.46 & $-1.44$ & $-1.22$ & 13.06 & 1.00 & 0 & $-18.20$  &\cr
  35 & NGC 4497 & S0/a & 12 31 32.56 & 11 37 29.3 & 1123 & 15.61 & $-1.04$ & $-0.86$ & 13.19 & 1.00 & 0 & $-18.14$  &\cr
  36 & NGC 4344 & BCD  & 12 23 37.76 & 17 32 27.0 & 1142 & 15.79 & $-1.48$ & $-1.12$ & 13.34 & 1.00 & 0 & $-17.91$  &\cr
  37 & NGC 4336 & S0/a & 12 23 29.85 & 19 25 36.9 & 1031 & 15.64 & $-1.03$ & $-1.03$ & 13.48 & 1.00 & 0 & $-17.84$  &\cr
  38 & NGC 4607 & Sb   & 12 41 12.26 & 11 53 06.5 & 2257 & 16.84 & $-1.19$ & $-1.06$ & 13.75 & 1.00 & 0 & $-17.54$  &\cr
  39 & IC 3475  & dI   & 12 32 40.65 & 12 46 10.0 & 2583 & 18.34 & $-1.28$ & $-1.34$ & 13.82 & 1.00 & 0 & $-17.44$  &\cr
  40 & NGC 4328 & S0   & 12 23 20.05 & 15 49 13.5 &  499 & 16.49 & $-1.28$ & $-0.92$ & 14.04 & 1.00 & 0 & $-17.22$  &\cr
  41 & IC 3718  & dI   & 12 44 45.39 & 12 21 03.2 &  849 & 16.61 & $-1.23$ & $-1.05$ & 14.07 & 1.00 & 0 & $-17.21$  &\cr
  42 & IC 3499  & S0/a & 12 33 45.02 & 10 59 45.0 & 1212 & 15.52 & $-1.10$ & $-0.65$ & 14.12 & 1.00 & 0 & $-17.17$  &\cr
  43 & IC 3470  & dE,N & 12 32 23.42 & 11 15 47.0 & 1500 & 16.00 & $-1.12$ & $-0.76$ & 14.29 & 1.00 & 0 & $-17.09$  &\cr
  44 & NGC 4641 & S0   & 12 43 07.64 & 12 03 03.4 & 2017 & 15.80 & $-1.00$ & $-0.52$ & 14.23 & 1.00 & 0 & $-17.06$  &\cr
  45 & NGC 4640 & dS0,N& 12 42 57.69 & 12 17 12.6 & 1931 & 16.80 & $-1.17$ & $-0.98$ & 14.37 & 1.00 & 0 & $-16.92$  &\cr 
  46 & IC 810   & S0   & 12 42 09.06 & 12 35 48.6 & $-$169 & 15.78 & $ -0.93$ & $-0.65$ & 14.41 & 1.00 & 0 & $-16.88$  &\cr
  47 & IC 809   & dE,N & 12 42 08.64 & 11 45 15.7 &  206 & 16.16 & $-1.10$ & $-0.78$ & 14.50 & 1.00 & 0 & $-16.78$  &\cr
  48 & IC 3727  & Scd  & 12 45 05.68 & 10 54 03.7 &   85 & 17.43 & $-1.32$ & $-1.05$ & 14.56 & 1.00 & 0 & $-16.70$  &\cr
  49 & IC 783   & S0/a & 12 21 38.81 & 15 44 42.5 & 1293 & 16.96 & $-1.20$ & $-1.08$ & 14.60 & 1.00 & 0 & $-16.65$  &\cr
  50 & IC 3652  & dE,N & 12 40 58.58 & 11 11 04.5 &  470 & 16.44 & $-1.24$ & $-0.87$ & 14.68 & 1.00 & 0 & $-16.60$  &\cr
  51 & IC 3457  & dE,N & 12 31 51.36 & 12 39 25.6 & 1263 & 17.57 & $-1.13$ & $-1.07$ & 14.69 & 1.00 & 0 & $-16.57$  &\cr
  52 &CGCG 098$-$132& E/S0 & 12 17 27.27 & 17 39 02.0 &  894 & 15.44 & $-0.95$ & $-0.50$ & 14.80 & 1.00 & 0 & $-16.50$  &\cr
  53 & IC 3653  & E    & 12 41 15.73 & 11 23 14.5 &  603 & 15.44 & $-0.89$ & $-0.55$ & 14.80 & 1.00 & 0 & $-16.48$  &\cr
  54 & NGC 4323 & S0   & 12 23 01.74 & 15 54 20.1 & 1803 & 16.97 & $-1.22$ & $-0.96$ & 14.81 & 1.00 & 0 & $-16.45$  &\cr
  55 & IC 3459  & dE,N & 12 31 55.99 & 12 10 26.9 &  278 & 17.76 & $-1.35$ & $-1.16$ & 14.83 & 1.00 & 0 & $-16.44$  &\cr
  56 & IC 3510  & dE,N & 12 34 14.85 & 11 04 17.8 & 1357 & 16.58 & $-1.19$ & $-0.80$ & 14.87 & 1.00 & 0 & $-16.41$  &\cr
  57 & IC 3540  & S0   & 12 35 27.21 & 12 45 00.8 &  753 & 15.89 & $-1.36$ & $-0.80$ & 14.94 & 1.00 & 0 & $-16.41$  &\cr
\noalign{\smallskip \hrule}
\noalign{\smallskip}\cr}}$$}
\end{table*}

\begin{table*}
{\vskip 0.55mm}
{$$\vbox{
\halign {\hfil #\hfil && \quad \hfil #\hfil \cr
\noalign{\hrule \medskip}
(1)  & (2) &  (3) & (4) & (5) & (6) &
(7)  & (8) & (9) & (10) &
(11) & (12) & (13) &\cr
ID  & Name &  Type & $\alpha$ (J2000) & $\delta$ (J2000) & $V_h$/km s$^{-1}$ &
$B(6)$   & ICP & OCP & $B_T$ &
$P$ & Class & $M_B$ &\cr
\noalign{\smallskip \hrule \smallskip}
  58 & IC 3720  & dE   & 12 44 47.44 & 12 03 53.4 &      & 18.53 & $-1.43$ & $-1.30$ & 14.98 & 1.00 & 1 & $-16.29$  &\cr
  59 & UGC 7346 & dE   & 12 18 41.78 & 17 43 07.5 &  819 & 17.25 & $-1.40$ & $-1.10$ & 15.03 & 1.00 & 0 & $-16.27$  &\cr 
  60 & UGC 7399A & dE,N & 12 20 48.83 & 17 29 14.0 & 1474 & 16.60 & $-1.13$ & $-0.91$ & 15.04 & 1.00 & 0 & $-16.24$  &\cr
  61**& VCC 723  & dS0  & 12 24 22.08 & 13 01 36.9 &  125 & 12.93 & $-1.32$ & $-0.39$ & 15.14 & 1.00 & 0 & $-16.19$  &\cr  
  62 & UGC 7436  & dE   & 12 22 19.57 & 14 45 39.5 &  923 & 16.55 & $-1.26$ & $-0.88$ & 15.11 & 1.00 & 0 & $-16.18$  &\cr 
  63 & VCC 459  & BCD  & 12 21 11.32 & 17 38 19.3 & 2107 & 15.92 & $-1.09$ & $-0.57$ & 15.18 & 1.00 & 0 & $-16.12$  &\cr 
  64 & VCC 1627 & E    & 12 35 37.25 & 12 22 54.9 &  236 & 15.91 & $-0.89$ & $-0.36$ & 15.41 & 1.00 & 0 & $-15.91$  &\cr
  65 & IC 3486  & dE,N & 12 33 14.04 & 12 51 27.8 & 1903 & 16.76 & $-1.22$ & at edge & 15.41 & 1.00 & 0 & $-15.87$  &\cr
  66 & UGC 7366 & dE,N & 12 19 28.66 & 17 13 50.1 &  925 & 16.59 & $-1.06$ & $-0.78$ & 15.40 & 1.00 & 0 & $-15.86$  &\cr 
  67 & IC 3578  & Scd  & 12 36 39.44 & 11 06 06.4 &  666 & 17.01 & $-1.14$ & $-0.93$ & 15.42 & 1.00 & 0 & $-15.86$  &\cr 
  68 & IC 3292  & dS0  & 12 24 48.38 & 18 11 42.6 &  710 & 16.39 & $-1.05$ & $-0.68$ & 15.43 & 1.00 & 0 & $-15.86$  &\cr 
  69 & IC 3298  & S    & 12 25 03.75 & 17 00 58.9 & 2452 & 16.65 & $-1.26$ & $-0.79$ & 15.43 & 1.00 & 0 & $-15.84$  &\cr 
  70 & IC 3313  & dE   & 12 25 36.45 & 15 49 47.6 & 1168 & 16.99 & $-1.30$ & $-0.91$ & 15.45 & 1.00 & 0 & $-15.83$  &\cr 
  71 & IC 3665  & dI   & 12 41 46.72 & 11 29 18.9 & 1227 & 17.62 & $-1.38$ & $-1.08$ & 15.44 & 1.00 & 0 & $-15.82$  &\cr 
  72 & IC 3586  & dS0  & 12 36 54.89 & 12 31 13.0 & 1547 & 16.54 & $-1.03$ & $-0.68$ & 15.58 & 1.00 & 0 & $-15.76$  &\cr 
  73 & LSBC F644$-$04 & dE,N & 12 25 36.21 & 15 50 51.4 &      & 18.21 & $-1.45$ & $-1.17$ & 15.56 & 0.84 & 1 & $-15.72$  &\cr 
  74 & IC 3461  & dE,N & 12 32 02.76 & 11 53 24.6 & 1038 & 16.70 & $-1.09$ & $-0.74$ & 15.61 & 1.00 & 0 & $-15.67$  &\cr 
  75 & VCC 1886 & dE,N & 12 41 39.44 & 12 14 51.2 & 1159 & 17.58 & $-1.37$ & $-1.02$ & 15.65 & 1.00 & 0 & $-15.64$  &\cr 
  76 & VCC 328  & dI   & 12 19 11.14 & 12 53 09.7 & 2179 & 18.42 & $-1.50$ & $-1.17$ & 15.77 & 1.00 & 0 & $-15.56$  &\cr 
  77 & IC 3388  & dE,N & 12 28 28.11 & 12 49 25.8 & 1704 & 17.12 & $-1.25$ & $-0.86$ & 15.72 & 1.00 & 0 & $-15.52$  &\cr 
  78 & IC 3779  & dE,N & 12 47 20.65 & 12 09 59.5 & 1193 & 16.99 & $-1.17$ & $-0.81$ & 15.74 & 1.00 & 0 & $-15.51$  &\cr 
  79 & VCC 1389 & dE,N & 12 31 52.04 & 12 28 54.9 &  936 & 17.51 & $-1.07$ & $-0.76$ & 15.78 & 1.00 & 0 & $-15.48$  &\cr 
  80 & IC 783 A & S0   & 12 22 19.64 & 15 44 01.0 & 1159 & 17.22 & $-1.20$ & $-0.85$ & 15.85 & 1.00 & 0 & $-15.42$  &\cr
  81 & VCC 1426 & dI   & 12 32 23.55 & 11 53 36.4 & 1110 & 18.46 & $-1.44$ & $-1.16$ & 15.96 & 1.00 & 0 & $-15.33$  &\cr 
  82 & IC 3509  & E    & 12 34 11.56 & 12 02 56.5 & 2050 & 17.04 & $-0.92$ & $-0.62$ & 16.01 & 1.00 & 0 & $-15.32$  &\cr  
  83 & IC 3443  & dE   & 12 31 15.77 & 12 19 54.9 & 1679 & 16.97 & $-0.95$ & $-0.70$ & 15.97 & 1.00 & 0 & $-15.28$  &\cr 
  84 & UGC 7425 & Scd  & 12 21 53.71 & 15 38 45.4 &  804 & 17.61 & $-1.22$ & $-0.94$ & 15.97 & 1.00 & 0 & $-15.28$  &\cr 
  85 & UGC 7504 & dI   & 12 25 21.63 & 16 25 47.0 &  913 & 17.29 & $-1.22$ & $-0.83$ & 15.98 & 1.00 & 0 & $-15.27$  &\cr 
 86* & VCC 841  & BCD  & 12 25 47.54 & 14 57 08.4 &  503 & 17.04 & $-1.10$ & $-0.70$ & 16.02 & 1.00 & 0 & $-15.25$  &\cr
  87 & VCC 1991 & dE,N & 12 44 09.40 & 11 10 35.8 &      & 18.04 & $-1.14$ & $-1.03$ & 16.09 & 0.62 & 1 & $-15.21$  &\cr
  88 & VCC 530  & dI   & 12 22 07.57 & 15 47 56.8 & 1299 & 18.70 & $-1.40$ & $-1.17$ & 16.07 & 1.00 & 0 & $-15.19$  &\cr 
  89 & IC 3466  & dI   & 12 32 05.71 & 11 49 04.2 &  903 & 17.07 & $-1.11$ & $-0.69$ & 16.10 & 1.00 & 0 & $-15.19$  &\cr  
  90 & VCC 1148 & E    & 12 28 58.13 & 12 39 40.2 & 1443 & 16.76 & $-0.68$ & at edge & 16.06 & 1.00 & 0 & $-15.18$  &\cr
  91 & VCC 1561 & dE,N & 12 34 25.03 & 12 54 15.6 &      & 19.41 & $-1.32$ & $-1.26$ & 16.16 & 0.95 & 1 & $-15.13$  &\cr 
  92 & VCC 753  & dE,N & 12 24 51.63 & 13 06 40.4 &      & 19.00 & $-1.34$ & $-1.20$ & 16.21 & 0.89 & 1 & $-15.09$  &\cr 
  93 & VCC 1185 & dE   & 12 29 23.55 & 12 27 03.4 &  500 & 17.71 & $-1.13$ & $-0.91$ & 16.18 & 1.00 & 0 & $-15.07$  &\cr  
  94 & IC 3490  & dE   & 12 33 13.94 & 10 55 43.0 &   80 & 17.81 & $-1.32$ & $-0.91$ & 16.27 & 1.00 & 0 & $-15.04$  &\cr  
  95 & VCC 1647 & dE   & 12 35 56.65 & 10 56 10.8 &      & 18.18 & $-1.37$ & $-1.00$ & 16.32 & 0.78 & 1 & $-14.94$  &\cr  
  96 & VCC 1982 & dE   & 12 43 51.07 & 11 28 01.3 &  938 & 17.57 & $-1.25$ & $-0.80$ & 16.34 & 1.00 & 0 & $-14.92$  &\cr  
  97 & VCC 1512 & S0   & 12 33 34.67 & 11 15 43.2 &  762 & 17.19 & $-0.91$ & $-0.57$ & 16.45 & 1.00 & 0 & $-14.92$  &\cr  
  98 & VCC 797  & dE,N & 12 25 24.09 & 18 08 23.6 &  773 & 17.64 & $-1.19$ & $-0.82$ & 16.37 & 1.00 & 0 & $-14.91$  &\cr  
  99 & VCC 1539 & dE,N & 12 34 06.75 & 12 44 30.0 & 1390 & 18.00 & $-1.02$ & $-0.93$ & 16.40 & 1.00 & 0 & $-14.89$  &\cr  
 100 & VCC 1921 & dS0  & 12 42 26.48 & 11 44 25.4 &      & 17.87 & $-1.24$ & $-0.88$ & 16.40 & 0.62 & 2 & $-14.88$  &\cr  
 101 & VCC 1711 & dE,N & 12 37 22.17 & 12 17 14.0 &      & 17.97 & $-1.23$ & $-0.90$ & 16.46 & 0.71 & 2 & $-14.87$  &\cr 
 102 & VCC 684  & dE,N & 12 23 57.74 & 12 53 14.0 &      & 17.92 & $-1.17$ & $-0.88$ & 16.47 & 0.62 & 2 & $-14.86$  &\cr 
 103 & VCC 1173 & dE   & 12 29 14.90 & 12 58 42.4 & 2468 & 17.60 & $-1.23$ & $-0.79$ & 16.39 & 1.00 & 0 & $-14.85$  &\cr 
 104 & VCC 1942 & dE,N & 12 42 50.75 & 12 18 30.6 &      & 18.47 & $-1.29$ & $-1.03$ & 16.50 & 0.77 & 2 & $-14.79$  &\cr 
 105 & UGC 7906 & dI   & 12 44 09.85 & 12 06 43.5 & 1010 & 18.58 & $-1.24$ & $-1.06$ & 16.51 & 1.00 & 0 & $-14.75$  &\cr
106* & VCC 815  & dE,N & 12 25 37.16 & 13 08 37.4 & $-$700 & 17.89 & $-1.26$ & $-0.84$ & 16.55 & 1.00 & 0 & $-14.73$  &\cr
107**& VCC 1995 & dE   & 12 44 16.79 & 12 01 38.7 &      & 18.15 & $-1.36$ & $-1.03$ & 16.56 & 0.71 & 1 & $-14.69$  &\cr 
 108 & VCC 846  & dE,N & 12 25 50.57 & 13 11 52.0 & $-$730 & 18.05 & $-1.30$ & $-0.88$ & 16.59 & 1.00 & 0 & $-14.68$  &\cr
 109 & VCC 618  & dI   & 12 23 07.44 & 13 44 40.3 & 1890 & 18.59 & $-1.34$ & $-1.01$ & 16.69 & 1.00 & 0 & $-14.67$  &\cr 
 110 &          & VLSB & 12 25 39.63 & 16 16 58.3 &      & 21.35 & $-1.28$ & $-1.39$ & 16.60 & 1.00 & 1 & $-14.66$  &\cr  
 111 & VCC 1717 & dE   & 12 37 28.98 & 12 21 08.9 &      & 19.27 & $-1.42$ & $-1.17$ & 16.67 & 0.91 & 1 & $-14.66$  &\cr 
 112 & UGC 7953 & dE,N & 12 47 16.29 & 11 45 38.9 &      & 18.08 & $-1.35$ & $-0.88$ & 16.61 & 0.67 & 2 & $-14.65$  &\cr  
113* & VCC 1870 & dE   & 12 41 15.35 & 11 17 54.8 &      & 17.80 & $-1.24$ & $-0.75$ & 16.74 & 0.45 & 2 & $-14.54$  &\cr 
 114 & LSBC F573$-$10 & dE   & 12 22 23.79 & 17 01 10.9 &      & 18.93 & $-1.50$ & at edge  & 16.73 &      & 1 & $-14.53$  &\cr
\noalign{\smallskip \hrule}
\noalign{\smallskip}\cr}}$$}
\end{table*}

\begin{table*}
{\vskip 0.55mm}
{$$\vbox{
\halign {\hfil #\hfil && \quad \hfil #\hfil \cr
\noalign{\hrule \medskip}
(1)  & (2) &  (3) & (4) & (5) & (6) &
(7)  & (8) & (9) & (10) &
(11) & (12) & (13) &\cr
ID  & Name &  Type & $\alpha$ (J2000) & $\delta$ (J2000) & $V_h$/km s$^{-1}$ &
$B(6)$  & ICP & OCP & $B_T$ &
$P$ & Class & $M_B$ &\cr
\noalign{\smallskip \hrule \smallskip}
 115 & VCC 1909 & dE,N & 12 42 07.42 & 11 49 42.0 &      & 18.00 & $-1.20$ & $-0.81$ & 16.76 & 0.56 & 2 & $-14.51$  &\cr
 116 & IC 3465  & dE,N & 12 32 12.29 & 12 03 41.8 & 1022 & 17.89 & $-1.14$ & $-0.74$ & 16.77 & 1.00 & 0 & $-14.51$  &\cr
 117 & VCC 1971 & dE   & 12 43 30.93 & 11 02 49.8 &      & 18.02 & $-1.27$ & $-0.80$ & 16.78 & 0.56 & 2 & $-14.51$  &\cr
 118 & VCC 1563 & dE,N & 12 34 26.09 & 11 55 01.2 &      & 19.37 & $-1.22$ & $-1.10$ & 16.82 & 0.89 & 1 & $-14.51$  &\cr
 119 & VCC 539  & dE,N & 12 22 14.82 & 14 08 31.5 &      & 18.77 & $-1.31$ & $-1.02$ & 16.84 & 0.87 & 2 & $-14.51$  &\cr
 120 & VCC 810  & dE,N & 12 25 33.58 & 13 13 38.3 & $-$340 & 18.34 & $-1.24$ & $-0.90$ & 16.78 & 1.00 & 0 & $-14.50$  &\cr
 121 & VCC 1149 & VLSB & 12 28 58.67 & 12 54 28.0 &      & 20.75 & $-1.35$ & $-1.33$ & 16.77 & 1.00 & 1 & $-14.47$  &\cr
 122 & VCC 2078 & dE   & 12 48 43.77 & 11 58 11.0 &      & 19.45 & $-1.19$ & $-1.17$ & 16.84 & 0.93 & 1 & $-14.45$  &\cr
 123 & VCC 878  & dE   & 12 26 10.13 & 14 55 44.9 &      & 19.36 & $-1.43$ & $-1.14$ & 16.87 & 0.87 & 1 & $-14.42$  &\cr
 124 & VCC 293  & dE,N & 12 18 31.89 & 13 11 28.4 &      & 18.84 & $-1.24$ & $-1.02$ & 16.91 & 0.89 & 1 & $-14.40$  &\cr
 125 & VCC 1625 & VLSB & 12 35 34.87 & 11 37 13.8 &      & 21.23 & $-1.21$ & $-1.36$ & 16.92 & 1.00 & 1 & $-14.40$  &\cr
 126 & VCC 1213 & dE,N & 12 29 39.28 & 12 32 54.1 &      & 18.40 & $-0.97$ & $-0.91$ & 16.88 & 0.79 & 2 & $-14.36$  &\cr
 127 & VCC 330  & dE,N & 12 19 12.47 & 12 51 07.2 &      & 18.66 & $-1.19$ & $-0.96$ & 16.97 & 0.88 & 2 & $-14.35$  &\cr
 128 & VCC 812  & dE,N & 12 25 34.86 & 15 11 40.0 &      & 18.52 & $-1.12$ & $-0.93$ & 16.96 & 0.85 & 2 & $-14.30$  &\cr
 129 & VCC 2081 & dE   & 12 49 46.25 & 11 13 32.4 &      & 19.07 & $-1.41$ & $-1.05$ & 17.02 & 0.88 & 1 & $-14.29$  &\cr
 130 & VCC 515  & dE   & 12 21 56.98 & 17 53 32.6 &      & 19.73 & $-1.42$ & $-1.18$ & 17.02 & 0.92 & 1 & $-14.28$  &\cr
131**& VCC 583  & dE   & 12 22 45.13 & 15 30 03.0 & $-$72 & 17.15 & $-2.15$ & $-0.62$ & 17.01 & 1.00 & 0 & $-14.26$  &\cr
 132 & VCC 1331 & dE/I & 12 30 58.60 & 11 42 28.2 &      & 20.28 & $-1.11$ & $-1.26$ & 17.05 & 0.92 & 1 & $-14.23$  &\cr
 133 & VCC 1915 & dE   & 12 42 13.44 & 12 32 46.1 &      & 19.05 & $-1.49$ & $-1.04$ & 17.07 & 0.86 & 2 & $-14.22$  &\cr
 134 & VCC 1264 & dE,N & 12 30 10.94 & 12 11 44.0 &      & 18.92 & $-1.12$ & $-1.02$ & 17.04 & 0.89 & 1 & $-14.21$  &\cr
 135 & VCC 1663 & dE   & 12 36 27.12 & 11 53 20.5 &      & 20.34 & $-1.45$ & $-1.24$ & 17.15 & 0.92 & 1 & $-14.21$  &\cr
 136 & VCC 769  & dE   & 12 25 04.24 & 15 42 40.8 &      & 18.41 & $-1.32$ & $-0.84$ & 17.08 & 0.61 & 2 & $-14.18$  &\cr
 137 & VCC 1815 & dE   & 12 39 56.42 & 11 54 15.7 &      & 20.17 & $-1.35$ & $-1.23$ & 17.16 & 0.95 & 1 & $-14.13$  &\cr
138* & VCC 472  & dE   & 12 21 24.10 & 15 37 07.4 &      & 19.26 & $-1.40$ & $-1.07$ & 17.36 & 0.93 & 1 & $-14.12$  &\cr
 139 & VCC 1399 & dE,N & 12 32 00.80 & 12 37 13.2 &      & 18.34 & $-1.25$ & $-0.80$ & 17.15 & 0.51 & 2 & $-14.11$  &\cr
140**& VCC 725  & dE   & 12 24 24.53 & 15 04 33.6 &      & 17.31 & $-1.21$ & $-0.46$ & 17.15 & 0.02 & 1 & $-14.10$  &\cr
 141 & VCC 554  & dE,N & 12 22 24.34 & 15 28 15.8 &      & 19.33 & $-1.29$ & $-1.08$ & 17.17 & 0.92 & 1 & $-14.10$  &\cr
 142 & VCC 1879 & dE,N & 12 41 27.38 & 11 08 45.5 &      & 19.20 & $-1.22$ & $-1.04$ & 17.19 & 0.92 & 1 & $-14.09$  &\cr
 143 & VCC 1313 & BCD  & 12 30 48.52 & 12 02 42.1 & 1254 & 17.27 & $-0.71$ & $-0.11$ & 17.18 & 1.00 & 0 & $-14.09$  &\cr
 144 & VCC 793  & dI   & 12 25 21.31 & 13 04 14.4 & 1908 & 18.33 & $-1.34$ & $-0.89$ & 17.20 & 1.00 & 0 & $-14.09$  &\cr 
 145 & VCC 1396 & dE,N & 12 31 56.43 & 11 58 22.2 &      & 19.52 & $-1.17$ & $-1.12$ & 17.21 & 0.88 & 1 & $-14.06$  &\cr
 146 & VCC 594  & dE   & 12 22 51.14 & 15 16 30.7 &      & 19.36 & $-1.40$ & $-1.08$ & 17.23 & 0.92 & 1 & $-14.04$  &\cr
 147 & VCC 1565 & dE,N & 12 34 30.47 & 11 44 04.4 &      & 19.20 & $-1.10$ & $-1.09$ & 17.29 & 0.93 & 1 & $-14.03$  &\cr
 148 & VCC 1418 & dE,N & 12 32 11.39 & 12 30 25.4 &      & 19.02 & $-1.25$ & $-0.98$ & 17.25 & 0.90 & 2 & $-14.02$  &\cr
 149 & VCC 2011 & dE   & 12 45 04.23 & 12 21 03.8 &      & 19.26 & $-1.48$ & $-1.03$ & 17.27 & 0.92 & 1 & $-14.01$  &\cr
 150 & VCC 1123 & dE,N & 12 28 42.67 & 12 32 59.9 &      & 18.96 & $-1.05$ & $-0.96$ & 17.26 & 0.90 & 1 & $-13.99$  &\cr
 151 & VCC 663  & dE   & 12 23 42.92 & 18 39 43.1 &      & 18.90 & $-1.29$ & $-0.95$ & 17.30 & 0.87 & 2 & $-13.99$  &\cr
 152 &          & dI   & 12 51 06.88 & 12 03 39.4 &      & 18.83 & $-1.40$ & $-0.92$ & 17.31 & 0.87 & 2 & $-13.99$  &\cr
 153 & VCC 350  & dI   & 12 19 26.00 & 13 18 38.4 &  305 & 19.15 & $-1.32$ & $-0.99$ & 17.34 & 1.00 & 0 & $-13.99$  &\cr
 154 & VCC 1599 & dE   & 12 35 06.60 & 11 54 03.1 &      & 20.11 & $-1.46$ & $-1.19$ & 17.36 & 0.93 & 1 & $-13.99$  &\cr
 155 & VCC 1891 & dE,N & 12 41 48.94 & 11 11 29.5 &      & 18.32 & $-1.14$ & $-0.73$ & 17.30 & 0.43 & 2 & $-13.97$  &\cr
 156 & VCC 668  & dE   & 12 23 47.10 & 15 07 32.0 &      & 19.26 & $-1.38$ & $-1.03$ & 17.30 & 0.92 & 1 & $-13.95$  &\cr
 157 & VCC 1369 & dE,N & 12 31 33.39 & 12 03 49.8 &      & 18.50 & $-1.08$ & $-0.76$ & 17.33 & 0.43 & 2 & $-13.94$  &\cr
 158 & VCC 1551 & VLSB & 12 34 15.38 & 11 28 01.7 &      & 21.35 & $-1.44$ & $-1.33$ & 17.38 & 1.00 & 1 & $-13.93$  &\cr
 159 &          & dI   & 12 25 46.29 & 16 38 07.8 &      & 18.54 & $-1.10$ & $-0.79$ & 17.33 & 0.51 & 2 & $-13.92$  &\cr
 160 & VCC 748  & dE   & 12 24 47.59 & 14 34 35.6 &      & 19.11 & $-1.36$ & $-0.96$ & 17.39 & 0.91 & 2 & $-13.90$  &\cr
 161 &          & VLSB & 12 32 34.62 & 12 38 15.4 &      & 21.10 & $-1.67$ & $-1.45$ & 17.38 & 1.00 & 1 & $-13.89$  &\cr
 162 & VCC 872  & dE,N & 12 26 06.72 & 12 51 40.0 & 1265 & 18.47 & $-0.97$ & $-0.73$ & 17.41 & 1.00 & 0 & $-13.87$  &\cr
163**& VCC 802  & BCD  & 12 25 28.72 & 13 29 51.5 & $-$215 & 18.34 & $-0.90$ & $-0.60$ & 17.45 & 1.00 & 0 & $-13.87$  &\cr
164* & VCC 422  &  dE  & 12 20 30.13 & 18 19 16.0 &      & 19.68 & $-1.43$ & $-1.10$ & 17.46 & 0.91 & 1 & $-13.85$  &\cr
 165 & VCC 1352 & dE   & 12 31 19.60 & 12 36 41.9 &      & 18.63 & $-1.19$ & $-0.80$ & 17.42 & 0.54 & 2 & $-13.83$  &\cr
 166 & VCC 1858 & dE   & 12 40 54.17 & 12 31 56.3 &      & 20.29 & $-1.32$ & $-1.20$ & 17.48 & 0.92 & 1 & $-13.83$  &\cr
 167 & VCC 1951 & dE,N & 12 43 02.26 & 11 41 53.3 &      & 19.03 & $-1.03$ & $-0.93$ & 17.45 & 0.91 & 1 & $-13.82$  &\cr
 168 & VCC 2088 & dE,N & 12 51 11.12 & 11 14 39.0 &      & 19.85 & $-1.05$ & $-1.14$ & 17.49 & 0.91 & 1 & $-13.82$  &\cr
 169 & VCC 761  & dE   & 12 25 00.28 & 15 36 15.8 &      & 19.36 & $-1.30$ & $-1.02$ & 17.44 & 0.93 & 1 & $-13.81$  &\cr
 170 & VCC 1606 & dE,N & 12 35 14.74 & 12 14 15.3 &      & 18.90 & $-0.90$ & $-0.86$ & 17.51 & 0.76 & 2 & $-13.80$  &\cr
 171 &          & dE   & 12 31 36.86 & 11 00 28.9 &      & 19.29 & $-1.33$ & $-0.99$ & 17.50 & 0.93 & 1 & $-13.79$  &\cr
\noalign{\smallskip \hrule}
\noalign{\smallskip}\cr}}$$}
\end{table*}

\begin{table*}
{\vskip 0.55mm}
{$$\vbox{
\halign {\hfil #\hfil && \quad \hfil #\hfil \cr
\noalign{\hrule \medskip}
(1)  & (2) &  (3) & (4) & (5) & (6) &
(7)  & (8) & (9) & (10) &
(11) & (12) & (13) &\cr
ID  & Name &  Type & $\alpha$ (J2000) & $\delta$ (J2000) & $V_h$/km s$^{-1}$ &
$B(6)$  & ICP & OCP & $B_T$ &
$P$ & Class & $M_B$ &\cr
\noalign{\smallskip \hrule \smallskip}
 172 & VCC 1366 & dE,N & 12 31 31.72 & 11 36 11.3 &      & 19.31 & $-0.96$ & $-0.98$ & 17.56 & 0.93 & 1 & $-13.78$  &\cr
 173 & VCC 2032 & dE   & 12 45 52.39 & 11 14 45.9 &      & 19.91 & $-1.25$ & $-1.13$ & 17.52 & 0.95 & 1 & $-13.77$  &\cr
 174 & VCC 833  & dE,N & 12 25 44.66 & 13 01 20.0 &  720 & 18.81 & $-1.00$ & $-0.82$ & 17.53 & 1.00 & 0 & $-13.75$  &\cr
 175 & VCC 1798 & VLSB & 12 39 31.27 & 11 27 14.3 &      & 21.57 & $-1.42$ & $-1.34$ & 17.56 & 1.00 & 1 & $-13.74$  &\cr
 176 & VCC 646  & dE   & 12 23 31.80 & 17 47 40.5 &      & 19.09 & $-1.27$ & $-0.91$ & 17.55 & 0.89 & 2 & $-13.73$  &\cr
 177 &          & VLSB & 12 44 22.28 & 12 00 34.1 &      & 21.35 & $-1.03$ & $-1.34$ & 17.55 & 1.00 & 1 & $-13.70$  &\cr
178**& VCC 1689 & dE,N & 12 36 51.32 & 12 22 08.7 &      & 19.49 & $-1.43$ & $-1.34$ & 17.65 & 1.00 & 1 & $-13.69$  &\cr
 179 & VCC 2062 & dI   & 12 48 00.03 & 10 58 14.7 & 1140 & 20.07 & $-1.40$ & $-0.91$ & 17.64 & 1.00 & 0 & $-13.68$  &\cr
 180 &          & VLSB & 12 20 51.26 & 16 21 49.6 &      & 21.32 & $-1.47$ & $-1.31$ & 17.60 & 1.00 & 1 & $-13.67$  &\cr
 181 & VCC 1905 & dE   & 12 42 03.13 & 12 28 50.8 &      & 20.52 & $-1.42$ & $-1.21$ & 17.65 & 0.90 & 1 & $-13.65$  &\cr
 182 & VCC 1464 & dE   & 12 32 53.87 & 11 11 28.8 &      & 19.51 & $-1.40$ & $-0.99$ & 17.72 & 0.95 & 1 & $-13.64$  &\cr
183**& VCC 1403 & dE/I & 12 32 00.37 & 13 04 58.3 &      & 19.80 & $-1.57$ & $-3.84$ & 17.64 & 1.00 & 1 & $-13.63$  &\cr
 184 & VCC 1681 & dE,N & 12 36 37.44 & 11 09 13.1 &      & 19.72 & $-1.13$ & $-1.05$ & 17.67 & 0.93 & 1 & $-13.62$  &\cr
 185 & VCC 863  & dE   & 12 25 59.75 & 14 02 22.5 &      & 20.37 & $-1.24$ & $-1.17$ & 17.74 & 0.93 & 1 & $-13.62$  &\cr
186**& VCC 1683 & dE   & 12 36 38.46 & 10 56 15.9 &      & 15.33 & $-2.61$ & $-0.42$ & 17.65 & 0.46 & 1 & $-13.60$  &\cr
 187 &          & VLSB & 12 23 22.16 & 19 12 11.5 &      & 22.01 & $-1.51$ & $-1.36$ & 17.75 & 1.00 & 1 & $-13.58$  &\cr
 188 & VCC 677  & dE   & 12 23 53.34 & 18 37 56.8 &      & 19.04 & $-1.33$ & $-0.85$ & 17.72 & 0.71 & 2 & $-13.56$  &\cr
 189 & VCC 696  & dE   & 12 24 04.04 & 17 32 57.5 &      & 19.15 & $-1.36$ & $-0.89$ & 17.72 & 0.79 & 2 & $-13.55$  &\cr
 190 & VCC 1191 & dE,N & 12 29 28.72 & 12 29 47.0 &      & 19.11 & $-1.11$ & $-0.87$ & 17.71 & 0.74 & 2 & $-13.54$  &\cr
 191 & VCC 795  & dE,N & 12 25 23.18 & 14 48 12.9 &      & 18.81 & $-1.18$ & $-0.72$ & 17.77 & 0.57 & 2 & $-13.51$  &\cr
192* & VCC 625  & dE   & 12 23 11.25 & 14 51 45.0 &      & 19.17 & $-1.26$ & $-0.85$ & 17.79 & 0.69 & 2 & $-13.50$  &\cr
 193 & VCC 818  & dI   & 12 25 37.79 & 16 39 51.8 &      & 19.59 & $-1.38$ & $-0.98$ & 17.80 & 0.95 & 1 & $-13.45$  &\cr
 194 & VCC 813  & dE,N & 12 25 35.75 & 16 35 46.1 &      & 19.81 & $-0.98$ & $-1.04$ & 17.80 & 0.95 & 1 & $-13.45$  &\cr
195* & VCC 1517 & dE,N & 12 33 40.85 & 12 34 17.0 &      & 19.49 & $-1.11$ & $-0.95$ & 17.82 & 0.97 & 1 & $-13.45$  &\cr
196**& VCC 716  & dE   & 12 24 13.70 & 14 55 44.6 &      & 18.97 & $-1.31$ & $-1.73$ & 17.84 & 1.00 & 1 & $-13.42$  &\cr
 197 & VCC 877  & dE,N & 12 26 09.58 & 13 40 23.7 &      & 19.04 & $-1.23$ & $-0.77$ & 17.89 & 0.64 & 2 & $-13.42$  &\cr
198* & VCC 779  & dE,N & 12 25 13.15 & 13 01 32.0 &      & 19.55 & $-0.97$ & $-0.95$ & 18.09 & 0.98 & 1 & $-13.42$  &\cr
 199 &          & dE   & 12 23 05.16 & 15 55 54.5 &      & 19.60 & $-1.38$ & $-0.98$ & 17.85 & 0.95 & 1 & $-13.41$  &\cr
200**& VCC 454  & dE   & 12 21 05.45 & 15 43 13.4 &      & 19.71 & $-1.44$ & $-1.18$ & 17.84 & 0.93 & 1 & $-13.41$  &\cr
 201 & VCC 1609 & dE,N & 12 35 20.34 & 11 38 10.6 &      & 19.16 & $-0.92$ & $-0.77$ & 17.91 & 0.63 & 1 & $-13.41$  &\cr
 202 & VCC 861  & dE   & 12 25 58.93 & 15 16 37.7 &      & 19.08 & $-1.31$ & $-0.80$ & 17.85 & 0.63 & 2 & $-13.40$  &\cr
 203 & VCC 1642 & dE,N & 12 35 53.17 & 11 40 55.4 &      & 19.52 & $-0.78$ & $-0.93$ & 17.91 & 0.97 & 1 & $-13.40$  &\cr
 204 & VCC 1413 & dE   & 12 32 07.71 & 12 26 03.0 &      & 19.80 & $-1.29$ & $-1.02$ & 17.89 & 0.95 & 1 & $-13.38$  &\cr
 205 & VCC 1153 & dE   & 12 28 59.85 & 12 38 55.0 &      & 19.09 & $-1.28$ & $-0.80$ & 17.88 & 0.63 & 2 & $-13.36$  &\cr
 206 & VCC 1216 & dE   & 12 29 41.36 & 12 02 47.5 &      & 20.41 & $-1.26$ & $-1.14$ & 17.92 & 0.89 & 1 & $-13.35$  &\cr
 207 &          & dE   & 12 17 32.66 & 18 24 17.3 &      & 19.76 & $-1.52$ & $-0.99$ & 17.96 & 0.95 & 1 & $-13.32$  &\cr
 208 & VCC 650  & dE,N & 12 23 34.10 & 13 17 59.1 &      & 19.32 & $-0.71$ & $-0.79$ & 18.14 & 0.78 & 1 & $-13.30$  &\cr
 209 & VCC 674  & dE,N & 12 23 52.68 & 13 52 56.9 &      & 19.25 & $-0.74$ & $-0.79$ & 18.04 & 0.71 & 1 & $-13.28$  &\cr
 210 & VCC 2001 & dE   & 12 44 33.77 & 11 47 42.9 &      & 19.67 & $-1.34$ & $-0.95$ & 18.00 & 0.98 & 1 & $-13.26$  &\cr
 211 & VCC 777  & dE   & 12 25 11.34 & 14 26 29.1 &      & 19.82 & $-1.33$ & $-0.98$ & 18.05 & 0.97 & 1 & $-13.26$  &\cr
 212 & VCC 1259 & dE   & 12 30 06.14 & 12 22 38.1 &      & 19.70 & $-1.40$ & $-0.96$ & 18.01 & 0.97 & 1 & $-13.25$  &\cr
 213 & VCC 1794 & dE,N & 12 39 27.07 & 11 46 35.3 &      & 18.99 & $-1.15$ & $-0.64$ & 18.08 & 0.66 & 2 & $-13.24$  &\cr
 214 & VCC 1136 & dE,N & 12 28 49.05 & 12 07 57.7 &      & 20.31 & $-1.09$ & $-1.10$ & 18.04 & 0.94 & 1 & $-13.23$  &\cr
 215 &          & dI   & 12 24 45.54 & 13 20 25.6 &      & 19.42 & $-1.22$ & $-0.82$ & 18.11 & 0.77 & 2 & $-13.22$  &\cr
 216 & VCC 1783 & dE   & 12 39 13.50 & 12 06 03.0 &      & 19.78 & $-1.07$ & $-0.94$ & 18.14 & 0.97 & 1 & $-13.22$  &\cr
 217 & VCC 726  & dE,N & 12 24 24.14 & 16 23 13.8 &      & 20.42 & $ -.96$ & $-1.12$ & 18.06 & 0.93 & 1 & $-13.20$  &\cr
218* & VCC 378  & dE   & 12 19 50.57 & 15 40 17.1 &      & 20.24 & $-1.41$ & $-1.09$ & 18.05 & 0.97 & 1 & $-13.20$  &\cr
 219 & VCC 1131 & dE   & 12 28 45.81 & 12 01 18.8 &      & 19.97 & $-1.44$ & $-1.01$ & 18.07 & 0.98 & 1 & $-13.20$  &\cr
 220 & VCC 1785 & dE,N & 12 39 15.73 & 11 16 11.7 &      & 19.54 & $-1.01$ & $-0.88$ & 18.09 & 0.84 & 1 & $-13.20$  &\cr
 221 & VCC 1831 & dE,N & 12 40 18.34 & 10 59 47.7 &      & 19.93 & $-1.17$ & $-1.00$ & 18.09 & 0.98 & 1 & $-13.16$  &\cr
 222 & VCC 2003 & dE   & 12 44 41.47 & 11 31 13.6 &      & 19.90 & $-1.21$ & $-0.96$ & 18.11 & 1.00 & 1 & $-13.16$  &\cr
 223 &          & VLSB & 12 36 35.18 & 11 37 03.9 &      & 21.32 & $-1.02$ & $-1.25$ & 18.15 & 1.00 & 1 & $-13.16$  &\cr
 224 & VCC 1454 & dE,N & 12 32 44.95 & 10 56 58.5 &      & 20.31 & $-1.08$ & $-1.07$ & 18.18 & 0.97 & 1 & $-13.15$  &\cr
 225 &          & dI   & 12 41 12.07 & 10 55 59.3 &      & 19.15 & $-1.12$ & $-0.69$ & 18.17 & 0.61 & 2 & $-13.08$  &\cr
 226 & VCC 780  & dE   & 12 25 13.24 & 14 50 44.7 &      & 19.77 & $-1.30$ & $-0.93$ & 18.19 & 0.97 & 1 & $-13.08$  &\cr
 227*&          & VLSB & 12 25 22.24 & 19 40 28.9 &      & 21.38 & $-1.13$ & $-1.21$ & 18.22 & 0.90 & 2 & $-13.08$  &\cr
 228 &          & dE   & 12 22 59.49 & 16 58 58.7 &      & 18.95 & $-1.15$ & $-0.58$ & 18.20 & 0.45 & 2 & $-13.06$  &\cr
\noalign{\smallskip \hrule}
\noalign{\smallskip}\cr}}$$}
\end{table*}

\begin{table*}
{\vskip 0.55mm}
{$$\vbox{
\halign {\hfil #\hfil && \quad \hfil #\hfil \cr
\noalign{\hrule \medskip}
(1)  & (2) &  (3) & (4) & (5) & (6) &
(7)  & (8) & (9) & (10) &
(11) & (12) & (13) &\cr
ID  & Name &  Type & $\alpha$ (J2000) & $\delta$ (J2000) & $V_h$/km s$^{-1}$ &
$B(6)$  & ICP & OCP & $B_T$ &
$P$ & Class & $M_B$ &\cr
\noalign{\smallskip \hrule \smallskip}
 229 &          & dE   & 12 24 36.57 & 18 56 15.2 &      & 20.78 & $-1.30$ & $-1.12$ & 18.30 & 0.93 & 1 & $-13.00$  &\cr
 230 &          & dE   & 12 19 23.32 & 19 35 14.2 &      & 19.34 & $-1.28$ & $-0.73$ & 18.28 & 0.75 & 2 & $-12.98$  &\cr
 231 & VCC 1736 & dE   & 12 37 55.18 & 11 08 56.3 &      & 20.78 & $-1.40$ & $-1.14$ & 18.33 & 0.90 & 1 & $-12.94$  &\cr
 232 & VCC 1278 & dE   & 12 30 17.50 & 12 14 28.3 &      & 20.36 & $-1.41$ & $-1.04$ & 18.32 & 0.98 & 1 & $-12.93$  &\cr
 233 & VCC 1815 & dE   & 12 39 57.72 & 11 54 26.0 &      & 21.07 & $-1.38$ & at edge & 18.38 &      & 1 & $-12.92$  &\cr
234**& VCC 635  & dE   & 12 23 21.99 & 13 20 38.4 &      & 19.43 & $-1.21$ & $-1.21$ & 18.56 & 0.96 & 2 & $-12.89$  &\cr
 235 & VCC 1637 & dE,N & 12 35 45.56 & 12 10 53.3 &      & 19.44 & $-0.99$ & $-0.69$ & 18.45 & 0.78 & 2 & $-12.88$  &\cr
 236 & VCC 714  & dE   & 12 24 13.49 & 17 30 16.1 &      & 19.41 & $-1.29$ & $-0.70$ & 18.41 & 0.80 & 2 & $-12.86$  &\cr
237* &          & VLSB & 12 44 54.57 & 10 57 44.6 &      & 21.49 & $-1.30$ & $-1.24$ & 18.42 & 0.84 & 1 & $-12.85$  &\cr
 238 & VCC 1880 & dE   & 12 41 28.03 & 12 25 40.4 &      & 19.57 & $-1.31$ & $-0.75$ & 18.45 & 0.78 & 1 & $-12.85$  &\cr
 239 & VCC 757  & dE   & 12 24 53.10 & 14 39 19.1 &      & 19.51 & $-1.36$ & $-0.71$ & 18.47 & 0.77 & 2 & $-12.84$  &\cr
240* & VCC 1700 & dE   & 12 37 03.30 & 11 28 42.9 &      & 21.14 & $-1.34$ & $-1.17$ & 18.48 & 0.91 & 1 & $-12.83$  &\cr
241**& VCC 505  & dE,N & 12 21 48.99 & 18 25 45.0 &      & 20.35 & $-1.41$ & $-1.41$ & 18.49 & 1.00 & 1 & $-12.83$  &\cr
 242 & VCC 1754 & dE   & 12 38 17.12 & 11 10 51.1 &      & 20.67 & $-1.40$ & $-1.09$ & 18.45 & 0.97 & 1 & $-12.82$  &\cr
 243 & VCC 1402 & dE,N & 12 32 00.28 & 11 01 24.1 &      & 19.86 & $-1.25$ & $-0.85$ & 18.50 & 0.88 & 1 & $-12.82$  &\cr
 244 &          & dE,N & 12 20 11.13 & 17 43 05.1 &      & 19.80 & $-1.33$ & $-0.81$ & 18.53 & 0.86 & 2 & $-12.78$  &\cr
 245 & VCC 521  & dE   & 12 22  2.84 & 17 12 03.8 &      & 20.09 & $-1.40$ & $-0.92$ & 18.53 & 0.97 & 1 & $-12.73$  &\cr
 246 &          & dE   & 12 41 05.03 & 12 15 58.1 &      & 19.76 & $-1.27$ & $-0.78$ & 18.58 & 0.87 & 1 & $-12.73$  &\cr
 247*& VCC 850  & dI   & 12 25 52.79 & 13 11 32.7 &      & 19.97 & $-1.28$ & $-1.00$ & 18.56 & 1.00 & 2 & $-12.71$  &\cr
 248 & VCC 644  & dE   & 12 23 28.89 & 17 32 25.0 &      & 19.73 & $-1.35$ & $-0.75$ & 18.58 & 0.86 & 2 & $-12.71$  &\cr
 249 & VCC 1595 & dE   & 12 34 59.99 & 11 32 50.0 &      & 19.73 & $-1.26$ & $-0.74$ & 18.64 & 0.86 & 1 & $-12.71$  &\cr
 250 & VCC 2025 & dE   & 12 45 35.37 & 11 33 07.5 &      & 20.02 & $-1.23$ & $-0.82$ & 18.58 & 0.88 & 1 & $-12.68$  &\cr
 251 & VCC 742  & dE,N & 12 24 40.40 & 15 42 35.9 &      & 20.12 & $-0.83$ & $-0.91$ & 18.58 & 0.93 & 1 & $-12.68$  &\cr
 252 & VCC 1613 & dE   & 12 35 26.88 & 12 31 41.3 &      & 19.81 & $-1.27$ & $-0.77$ & 18.64 & 0.87 & 1 & $-12.68$  &\cr
253**& VCC 624  & dE,N & 12 23 11.66 & 13 25 06.8 &      & 20.55 & $-1.01$ & $-0.99$ & 18.75 & 1.00 & 2 & $-12.67$  &\cr
254* & VCC 519  & dE   & 12 22 00.87 & 14 08 11.6 &      & 20.52 & $-1.37$ & $-1.00$ & 18.70 & 1.00 & 1 & $-12.66$  &\cr
255**& VCC 721  & dI   & 12 24 22.02 & 13 25 04.6 &      & 20.01 & $-0.97$ & $-0.83$ & 18.69 & 0.90 & 2 & $-12.65$  &\cr
 256 &          & dE,N & 12 35 59.74 & 11 27 08.3 &      & 20.39 & $-0.94$ & $-0.97$ & 18.66 & 1.00 & 2 & $-12.65$  &\cr
257* & VCC 1594 & dE   & 12 35 00.08 & 11 20 38.3 &      & 19.58 & $-1.23$ & $-0.67$ & 18.65 & 0.81 & 2 & $-12.64$  &\cr
 258 & VCC 1986 & dE,N & 12 43 57.80 & 11 52 50.7 &      & 20.24 & $-0.92$ & $-0.94$ & 18.61 & 0.98 & 1 & $-12.64$  &\cr
 259 & VCC 1963 & dE,N & 12 43 18.07 & 11 28 30.6 &      & 20.88 & $-0.70$ & $-1.10$ & 18.64 & 0.97 & 1 & $-12.62$  &\cr
 260 & VCC 789  & dE   & 12 25 19.16 & 13 15 24.0 &      & 20.36 & $-1.33$ & $-0.96$ & 18.68 & 1.00 & 1 & $-12.62$  &\cr
 261 & VCC 1518 & dE   & 12 33 40.89 & 12 22 56.7 &      & 20.16 & $-1.33$ & $-0.90$ & 18.67 & 0.93 & 1 & $-12.61$  &\cr
 262 & VCC 1729 & dE   & 12 37 46.06 & 10 59 07.1 &      & 19.79 & $-1.30$ & $-0.77$ & 18.65 & 0.86 & 2 & $-12.60$  &\cr
 263 & VCC 1522 & dE,N & 12 33 47.08 & 11 46 54.4 &      & 20.83 & $-0.89$ & $-1.06$ & 18.75 & 0.96 & 1 & $-12.57$  &\cr
 264 & VCC 1718 & dE,N & 12 37 30.13 & 11 28 54.4 &      & 19.83 & $-1.10$ & $-0.74$ & 18.74 & 0.87 & 1 & $-12.57$  &\cr
 265 & VCC 1405 & dE   & 12 32 00.05 & 11 18 06.9 &      & 21.19 & $-1.33$ & $-1.12$ & 18.81 & 0.92 & 1 & $-12.56$  &\cr
 266 &          & VLSB & 12 21 52.38 & 17 29 58.7 &      & 21.59 & $-1.39$ & $-1.21$ & 18.73 & 0.89 & 2 & $-12.55$  &\cr
 267 &          & dE   & 12 38 38.84 & 11 28 53.8 &      & 19.35 & $-1.13$ & $-0.48$ & 18.77 & 0.51 & 2 & $-12.51$  &\cr
 268 & VCC 708  & dE,N & 12 24 13.41 & 13 37 57.5 &      & 19.86 & $-1.23$ & $-0.71$ & 18.83 & 0.87 & 2 & $-12.50$  &\cr
 269 & VCC 1536 & dE,N & 12 34 06.59 & 11 50 12.4 &      & 20.38 & $-0.69$ & $-0.92$ & 18.82 & 0.98 & 1 & $-12.49$  &\cr
270* & VCC 1381 & dE   & 12 31 44.05 & 12 36 45.2 &      & 20.12 & $-1.33$ & $-0.87$ & 18.78 & 0.92 & 1 & $-12.48$  &\cr
 271 &          & dI   & 12 30 24.03 & 12 26 09.5 &      & 21.31 & $-1.36$ & $-1.13$ & 18.76 & 0.95 & 1 & $-12.48$  &\cr
 272** & VCC 1746 & dE & 12 38 11.83 & 12 03 29.7 &      & 20.34 & $-1.50$ & $-1.93$ & 18.86 & 1.00 & 1 & $-12.46$  &\cr
 273 & VCC 845  & dE   & 12 25 48.16 & 13 51 15.1 &      & 20.16 & $-1.19$ & $-0.83$ & 18.86 & 0.90 & 2 & $-12.46$  &\cr
 274 & VCC 719  & dE   & 12 24 18.78 & 12 54 45.5 &      & 20.00 & $-1.23$ & $-0.75$ & 18.90 & 0.90 & 1 & $-12.43$  &\cr
 275 & VCC 557  & dE   & 12 22 29.14 & 13 18 56.4 &      & 19.67 & $-1.19$ & $-0.54$ & 18.98 & 0.56 & 2 & $-12.43$  &\cr
 276 & VCC 830  & dE   & 12 25 42.55 & 17 59 25.2 &      & 20.05 & $-1.25$ & $-0.59$ & 18.84 & 0.72 & 2 & $-12.42$  &\cr
 277 & VCC 643  & dE   & 12 23 29.45 & 14 53 22.2 &      & 20.01 & $-1.27$ & $-0.76$ & 18.86 & 0.89 & 1 & $-12.42$  &\cr
 278**&         & dE/I & 12 29 19.59 & 12 22 37.7 &      & 20.06 & $-1.20$ & $-3.14$ & 18.94 & 1.00 & 1 & $-12.41$  &\cr
 279 & VCC 2010 & dE   & 12 44 53.04 & 12 10 58.9 &      & 19.99 & $-1.29$ & $-0.75$ & 18.87 & 0.90 & 1 & $-12.40$  &\cr
 280 & VCC 432  & dE   & 12 20 46.73 & 17 13 52.0 &      & 19.63 & $-1.21$ & $-0.56$ & 18.89 & 0.57 & 2 & $-12.37$  &\cr
 281 &          & dE/I & 12 30 47.17 & 11 32 16.9 &      & 20.37 & $-1.41$ & $-0.87$ & 18.96 & 0.95 & 1 & $-12.37$  &\cr
 282 &          & VLSB & 12 51 01.10 & 11 29 01.4 &      & 21.66 & $-1.33$ & $-1.17$ & 19.01 & 0.91 & 1 & $-12.36$  &\cr
 283 & VCC 1621 & dE   & 12 35 34.08 & 11 47 10.1 &      & 19.92 & $-1.26$ & $-0.68$ & 18.97 & 0.91 & 2 & $-12.35$  &\cr
 284 & VCC 1925 & dE   & 12 42 33.77 & 11 49 14.6 &      & 21.43 & $-1.26$ & $-1.14$ & 18.95 & 0.95 & 1 & $-12.32$  &\cr
 285 &          & VLSB & 12 24 08.57 & 13 49 59.3 &      & 21.81 & $-1.28$ & $-1.20$ & 18.99 & 1.00 & 1 & $-12.32$  &\cr
\noalign{\smallskip \hrule}
\noalign{\smallskip}\cr}}$$}
\end{table*}

\begin{table*}
{\vskip 0.55mm}
{$$\vbox{
\halign {\hfil #\hfil && \quad \hfil #\hfil \cr
\noalign{\hrule \medskip}
(1)  & (2) &  (3) & (4) & (5) & (6) &
(7)  & (8) & (9) & (10) &
(11) & (12) & (13) &\cr
ID  & Name &  Type & $\alpha$ (J2000) & $\delta$ (J2000) & $V_h$/km s$^{-1}$ &
$B(6)$  & ICP & OCP & $B_T$ &
$P$ & Class & $M_B$ &\cr
\noalign{\smallskip \hrule \smallskip}
 286 &          & dE/I & 12 40 04.74 & 11 34 15.4 &      & 20.03 & $-1.21$ & $-0.69$ & 18.99 & 0.88 & 1 & $-12.31$  &\cr
 287 & VCC 1841 & dE   & 12 40 27.05 & 12 11 57.0 &      & 20.22 & $-1.31$ & $-0.80$ & 19.05 & 0.92 & 2 & $-12.29$  &\cr
 288 & IBM88 V03L12 & dE,N & 12 24 35.15 & 15 09 50.7 &  & 20.60 & $-1.31$ & $-0.94$ & 18.99 & 1.00 & 1 & $-12.28$ &\cr 
 289 &          & dE/I & 12 19 50.59 & 16 16 08.1 &      & 20.08 & $-1.22$ & $-0.74$ & 19.00 & 0.90 & 1 & $-12.27$  &\cr
 290   & VCC 1416 & dE,N & 12 32 10.30 & 12 33 02.6 &    & 21.26 & $-0.68$ & $-1.14$ & 19.00 & 0.96 & 1 & $-12.26$  &\cr
 291 & VCC 2072 & dE   & 12 48 25.23 & 12 14 15.7 &      & 20.52 & $-1.36$ & $-0.89$ & 19.03 & 0.97 & 1 & $-12.24$  &\cr
 292 &          & dE,N & 12 45 24.97 & 10 55 26.6 &      & 20.51 & $-0.82$ & $-0.89$ & 19.03 & 0.96 & 1 & $-12.23$  &\cr
 293 &          & dI   & 12 24 12.90 & 14 29 39.1 &      & 20.21 & $-1.32$ & $-0.76$ & 19.08 & 0.92 & 1 & $-12.22$  &\cr
 294*&          & dE/I & 12 32 07.37 & 11 20 31.5 &      & 21.08 & $-1.20$ & $-1.02$ & 19.16 & 0.97 & 1 & $-12.22$  &\cr
 295 & VCC 600  & dE   & 12 22 55.51 & 15 33 34.4 &      & 20.49 & $-1.14$ & $-0.87$ & 19.06 & 0.97 & 1 & $-12.20$  &\cr
296**& VCC 704  & dE   & 12 24 11.32 & 13 22 25.1 &      & 20.70 & $-1.21$ & $-1.28$ & 19.17 & 0.85 & 1 & $-12.18$  &\cr
 297 & VCC 1143 & dE,N & 12 28 55.69 & 12 42 29.3 &      & 20.05 & $-1.31$ & at edge & 19.07 &      & 1 & $-12.17$  &\cr
 298**&         & dE   & 12 40 14.56 & 11 31 57.2 &      & 20.28 & $-1.49$ & $-3.56$ & 19.15 & 1.00 & 1 & $-12.17$  &\cr
 299 & VCC 2047 & dE   & 12 47 08.59 & 11 27 39.9 &      & 19.83 & $-1.17$ & $-0.55$ & 19.12 & 0.67 & 1 & $-12.15$  &\cr
 300 & VCC 1161 & dE   & 12 29 05.45 & 12 01 52.9 &      & 20.25 & $-1.29$ & $-0.76$ & 19.11 & 0.95 & 1 & $-12.15$  &\cr
 301 & VCC 1904 & dE,N & 12 42 02.76 & 10 57 13.4 &      & 19.98 & $-0.43$ & $-0.64$ & 19.11 & 0.85 & 1 & $-12.14$  &\cr
 302 & VCC 754  & dE   & 12 24 50.72 & 15 00 38.9 &      & 20.23 & $-1.33$ & $-0.74$ & 19.11 & 0.92 & 1 & $-12.14$  &\cr
 303 & VCC 536  & dE   & 12 22 12.26 & 16 58 28.0 &      & 20.51 & $-1.33$ & $-0.86$ & 19.12 & 0.97 & 1 & $-12.14$  &\cr
 304 &          & dE/I & 12 25 55.13 & 19 11 51.1 &      & 20.51 & $-1.09$ & $-0.86$ & 19.16 & 0.97 & 2 & $-12.14$  &\cr
 305 &          & dE   & 12 20 35.05 & 18 53 02.8 &      & 20.34 & $-1.31$ & $-0.76$ & 19.19 & 0.92 & 1 & $-12.12$  &\cr
 306*&          & dI   & 12 25 11.26 & 13 27 56.0 &      & 20.60 & $-1.32$ & $-0.86$ & 19.22 & 0.96 & 2 & $-12.11$  &\cr
307* & VCC 1990 & dE   & 12 44 06.81 & 12 41 03.9 &      & 20.94 & $-1.47$ & $-0.98$ & 19.17 & 1.00 & 1 & $-12.10$  &\cr
 308 & VCC 495  & dE   & 12 21 41.78 & 17 49 40.7 &      & 20.24 & $-1.25$ & $-0.67$ & 19.22 & 0.90 & 2 & $-12.08$  &\cr
 309   &        & VLSB & 12 37 37.71 & 12 28 38.5 &      & 22.23 & $-1.27$ & $-1.22$ & 19.27 & 7.00 & 1 & $-12.08$  &\cr
310**  &        & VLSB & 12 31 48.55 & 10 58 09.3 &      & 21.20 & $-1.01$ & $-1.96$ & 19.22 & 1.00 & 1 & $-12.08$  &\cr
 311 & VCC 547  & dE   & 12 22 20.91 & 15 09 35.2 &      & 20.29 & $-1.31$ & $-0.74$ & 19.21 & 0.90 & 1 & $-12.06$  &\cr
 312 & VCC 1271 & dE   & 12 30 15.31 & 12 30 58.1 &      & 20.50 & $-1.30$ & $-0.82$ & 19.20 & 0.94 & 1 & $-12.05$  &\cr
 313 &          & dI   & 12 25 20.95 & 13 49 09.4 &      & 21.04 & $-1.30$ & $-0.98$ & 19.27 & 1.00 & 1 & $-12.05$  &\cr
 314 & VCC 1157 & dE   & 12 29 02.05 & 12 26 05.5 &      & 20.73 & $-1.23$ & $-0.90$ & 19.21 & 0.96 & 1 & $-12.04$  &\cr
 315 &          & VLSB & 12 24 26.20 & 13 28 05.9 &      & 21.80 & $-1.32$ & $-1.11$ & 19.26 & 1.00 & 1 & $-12.04$  &\cr
 316 & VCC 829  & dE   & 12 25 42.78 & 15 34 30.1 &      & 20.14 & $-1.23$ & $-0.63$ & 19.22 & 0.82 & 2 & $-12.04$  &\cr
 317 & VCC 1461 & dE,N & 12 32 51.28 & 11 17 45.2 &      & 20.62 & $-1.13$ & $-0.81$ & 19.36 & 0.93 & 1 & $-12.03$  &\cr
 318 &          & VLSB & 12 22 09.57 & 15 39 10.7 &      & 21.27 & $-1.34$ & $-1.05$ & 19.24 & 1.00 & 1 & $-12.02$  &\cr
 319 & VCC 605  & dE,N & 12 23 02.05 & 13 33 33.1 &      & 20.60 & $-1.11$ & $-0.83$ & 19.39 & 0.95 & 2 & $-12.00$  &\cr
320**& VCC 844  & dE   & 12 25 48.40 & 13 07 21.5 &      & 20.33 & $-1.29$ & $-1.46$ & 19.28 & 1.00 & 1 & $-12.00$  &\cr
 321 &          & VLSB & 12 46 48.33 & 12 11 49.5 &      & 21.77 & $-1.24$ & $-1.15$ & 19.28 & 1.00 & 1 & $-11.99$  &\cr
 322 &          & dI   & 12 43 08.14 & 11 05 55.3 &      & 21.52 & $-1.19$ & $-1.09$ & 19.29 & 0.85 & 2 & $-11.99$  &\cr
 323 &          & dE,N & 12 34 01.41 & 12 43 11.1 &      & 20.67 & $-1.17$ & $-0.85$ & 19.33 & 0.95 & 1 & $-11.96$  &\cr
 324 &          & dE/I & 12 44 54.13 & 11 01 08.3 &      & 20.84 & $-1.34$ & $-0.90$ & 19.34 & 0.96 & 1 & $-11.96$  &\cr
 325 &          & dI   & 12 23 39.13 & 13 49 04.8 &      & 20.20 & $-1.30$ & $-0.61$ & 19.37 & 0.78 & 2 & $-11.95$  &\cr
 326 & VCC 1672 & dE,N & 12 36 32.63 & 12 31 05.9 &      & 20.23 & $-0.44$ & $-0.61$ & 19.41 & 7.73 & 1 & $-11.94$  &\cr
 327 & VCC 804  & dE,N & 12 25 30.69 & 12 58 37.9 &      & 21.03 & $-1.14$ & $-0.96$ & 19.35 & 1.00 & 1 & $-11.94$  &\cr
 328 &          & dE,N & 12 23 53.17 & 13 30 23.6 &      & 20.48 & $-0.95$ & $-0.72$ & 19.43 & 0.88 & 2 & $-11.93$  &\cr
 329 &          & dE,N & 12 33 51.17 & 12 57 30.8 &      & 20.34 & $-1.00$ & $-0.67$ & 19.41 & 0.87 & 1 & $-11.88$  &\cr
 330 & VCC 1466 & dE   & 12 32 55.37 & 12 38 06.9 &      & 20.21 & $-1.23$ & $-0.61$ & 19.40 & 0.73 & 2 & $-11.87$  &\cr
 331*&          & dE   & 12 35 19.70 & 11 29 20.8 &      & 21.00 & $-1.10$ & $-0.90$ & 19.50 & 0.94 & 1 & $-11.84$  &\cr
332* &          & dI   & 12 21 13.27 & 16 17 36.9 &      & 20.27 & $-1.05$ & $-0.64$ & 19.44 & 0.86 & 2 & $-11.83$  &\cr
 333 &          & dE   & 12 45 16.32 & 12 14 53.7 &      & 21.05 & $-1.21$ & $-0.84$ & 19.45 & 0.91 & 1 & $-11.82$  &\cr
 334 & VCC 647  & dE   & 12 23 33.43 & 17 49 14.9 &      & 20.73 & $-1.38$ & $-0.82$ & 19.46 & 0.94 & 1 & $-11.82$  &\cr
 335 & VCC 1558 & dI   & 12 34 21.35 & 11 24 57.7 &      & 20.42 & $-1.24$ & $-0.65$ & 19.51 & 0.88 & 2 & $-11.80$  &\cr
 336 & VCC 1538 & dE   & 12 34 05.98 & 11 03 15.7 &      & 20.74 & $-1.38$ & $-0.81$ & 19.49 & 0.94 & 1 & $-11.79$  &\cr
 337 &          & dE/I & 12 24 34.34 & 13 22 24.4 &      & 20.99 & $-1.43$ & $-0.88$ & 19.56 & 0.94 & 1 & $-11.79$  &\cr
 338 & VCC 603  & dE   & 12 22 59.28 & 13 45 25.0 &      & 20.40 & $-1.29$ & $-0.62$ & 19.57 & 0.80 & 2 & $-11.79$  &\cr
 339 &          & VLSB & 12 45 34.62 & 11 28 49.2 &      & 21.34 & $-1.32$ & $-1.00$ & 19.49 & 1.00 & 1 & $-11.77$  &\cr
 340 &          & dE/I & 12 48 00.30 & 11 22 21.5 &      & 20.95 & $-1.24$ & $-0.88$ & 19.51 & 0.96 & 1 & $-11.77$  &\cr
 341 &          & dE,N & 12 44 45.02 & 11 48 04.1 &      & 20.82 & $-0.80$ & $-0.84$ & 19.50 & 0.92 & 2 & $-11.76$  &\cr
 342 & IBM88 V03L15 & dE/I & 12 25 55.00 & 14 38 28.5 &  & 20.71 & $-1.26$ & $-0.77$ & 19.54 & 0.94 & 1 & $-11.75$ &\cr 
\noalign{\smallskip \hrule}
\noalign{\smallskip}\cr}}$$}
\end{table*}

\begin{table*}
{\vskip 0.55mm}
{$$\vbox{
\halign {\hfil #\hfil && \quad \hfil #\hfil \cr
\noalign{\hrule \medskip}
(1)  & (2) &  (3) & (4) & (5) & (6) &
(7)  & (8) & (9) & (10) &
(11) & (12) & (13) &\cr
ID  & Name &  Type & $\alpha$ (J2000) & $\delta$ (J2000) & $V_h$/km s$^{-1}$ &
$B(6)$  & ICP & OCP & $B_T$ &
$P$ & Class & $M_B$ &\cr
\noalign{\smallskip \hrule \smallskip}
 343 &          & VLSB & 12 43 17.11 & 11 37 41.4 &      & 21.46 & $-0.90$ & $-1.02$ & 19.53 & 1.00 & 1 & $-11.74$  &\cr
 344 & IBM88 V07L04& dI   & 12 29 53.69 & 12 37 13.1 &   & 21.45 & $-1.15$ & $-1.16$ & 19.51 & 1.00 & 1 & $-11.73$  &\cr
 345** & VCC 1103 & dE,N & 12 28 26.29 & 12 20 45.6 &    & 21.17 & $-0.71$ & $-1.00$ & 19.57 & 0.99 & 1 & $-11.70$  &\cr
346* &          & dE/I & 12 43 08.00 & 11 37 16.6 &      & 21.10 & $-1.41$ & $-0.91$ & 19.58 & 0.98 & 1 & $-11.69$  &\cr
 347 &          & dI   & 12 17 49.17 & 16 35 46.8 &      & 20.60 & $-1.28$ & $-0.68$ & 19.58 & 0.89 & 2 & $-11.68$  &\cr
 348 &          & dE/I & 12 26 08.19 & 15 28 53.6 &      & 20.37 & $-1.22$ & $-0.57$ & 19.61 & 0.73 & 2 & $-11.65$  &\cr
 349 & VCC 1277 & dE   & 12 30 18.02 & 12 02 30.4 &      & 20.83 & $-1.35$ & $-0.79$ & 19.63 & 0.94 & 1 & $-11.65$  &\cr
 350 & VCC 1968 & dE   & 12 43 27.26 & 10 57 23.0 &      & 20.34 & $-1.32$ & $-0.55$ & 19.63 & 0.70 & 1 & $-11.63$  &\cr
 351 &          & dE/I & 12 50 35.95 & 11 23 18.6 &      & 20.44 & $-1.41$ & $-0.61$ & 19.71 & 0.81 & 2 & $-11.62$  &\cr
 352 & VCC 1882 & dE,N & 12 41 30.80 & 11 40 56.2 &      & 20.81 & $-0.60$ & $-0.76$ & 19.68 & 0.90 & 2 & $-11.61$  &\cr
 353 & VCC 556  & dE   & 12 22 28.10 & 13 09 46.1 &      & 20.70 & $-1.20$ & $-0.68$ & 19.75 & 0.90 & 2 & $-11.61$  &\cr
 354 &          & dI   & 12 30 18.23 & 12 34 18.3 &      & 20.96 & $-1.31$ & $-0.83$ & 19.64 & 0.94 & 1 & $-11.60$  &\cr
 355 &          & dE/I & 12 31 36.45 & 13 05 19.4 &      & 20.99 & $-1.31$ & at edge & 19.67 &      & 1 & $-11.59$  &\cr
 356 & VCC 1500 & dE   & 12 33 22.56 & 11 38 29.9 &      & 20.40 & $-1.29$ & $-0.62$ & 19.78 & 0.79 & 2 & $-11.59$  &\cr
 357 &          & dE   & 12 44 52.86 & 11 33 34.1 &      & 20.69 & $-1.30$ & $-0.71$ & 19.68 & 0.88 & 1 & $-11.58$  &\cr
 358 &          & VLSB & 12 36 18.14 & 11 57 12.1 &      & 21.71 & $-1.48$ & $-1.01$ & 19.82 & 1.00 & 1 & $-11.56$  &\cr
 359 &          & dE/I & 12 50 45.45 & 11 54 45.5 &      & 20.84 & $-1.24$ & $-0.72$ & 19.77 & 0.86 & 1 & $-11.55$  &\cr
 360 &          & VLSB & 12 23 54.55 & 13 10 56.2 &      & 21.64 & $-1.14$ & $-0.99$ & 19.85 & 1.00 & 1 & $-11.54$  &\cr
 361 &          & dE,N & 12 42 22.81 & 11 41 07.2 &      & 21.34 & $-1.09$ & $-0.90$ & 19.76 & 0.98 & 1 & $-11.53$  &\cr
 362 & VCC 767  & dE   & 12 25 04.80 & 13 04 32.3 &      & 21.21 & $-1.26$ & $-0.88$ & 19.76 & 1.00 & 1 & $-11.53$  &\cr
 363 & VCC 1977 & dE   & 12 43 38.41 & 11 17 51.9 &      & 20.52 & $-1.26$ & $-0.58$ & 19.75 & 0.77 & 2 & $-11.52$  &\cr
 364 & VCC 1769 & dE   & 12 38 38.03 & 12 36 39.2 &      & 20.73 & $-1.35$ & $-0.67$ & 19.80 & 0.90 & 1 & $-11.52$  &\cr
 365 & VCC 1083 & dE   & 12 28 12.24 & 11 58 13.4 &      & 20.33 & $-1.22$ & $-0.46$ & 19.76 & 0.64 & 2 & $-11.51$  &\cr
 366 &          & dI   & 12 36 13.25 & 12 10 10.9 &      & 21.61 & $-1.22$ & $-0.98$ & 19.85 & 1.00 & 2 & $-11.50$  &\cr
 367 & VCC 1162 & dE,N & 12 29 05.19 & 12 09 14.0 &      & 20.93 & $-0.71$ & $-0.77$ & 19.78 & 0.92 & 2 & $-11.49$  &\cr
368**&          & VLSB & 12 23 11.67 & 15 23  9.4 &      & 21.40 & $-1.33$ & $-1.35$ & 19.79 & 1.00 & 1 & $-11.48$  &\cr
 369 &          & dE/I & 12 33 07.56 & 12 12 13.7 &      & 21.23 & $-1.02$ & $-0.87$ & 19.81 & 0.95 & 2 & $-11.48$  &\cr
 370 & VCC 1494 & dE   & 12 33 16.93 & 12 16 57.2 &      & 20.93 & $-1.43$ & $-0.73$ & 19.82 & 0.88 & 1 & $-11.46$  &\cr
 371 &          & dI   & 12 42 48.07 & 12 38 48.0 &      & 21.19 & $-1.11$ & $-0.85$ & 19.82 & 0.95 & 2 & $-11.46$  &\cr
 372 &          & dE/I & 12 22 19.99 & 15 40 46.8 &      & 20.53 & $-1.08$ & $-0.55$ & 19.82 & 0.72 & 2 & $-11.45$  &\cr
 373 & VCC 1578 & dE   & 12 34 41.76 & 11 08 34.2 &      & 20.67 & $-1.30$ & $-0.56$ & 19.88 & 0.77 & 2 & $-11.38$  &\cr
 374 &          & VLSB & 12 18 50.72 & 15 54 18.4 &      & 21.79 & $-1.37$ & $-1.01$ & 19.91 & 1.00 & 1 & $-11.36$  &\cr
 375 & VCC 1680 & dE   & 12 36 36.69 & 10 59 28.6 &      & 20.47 & $-1.18$ & $-0.46$ & 19.90 & 0.66 & 2 & $-11.35$  &\cr
 376 & VCC 607  & dE   & 12 23 02.28 & 13 54 50.1 &      & 20.91 & $-1.35$ & $-0.64$ & 20.00 & 0.87 & 1 & $-11.34$  &\cr
 377 &          & dI   & 12 23 49.02 & 15 14 39.8 &      & 20.94 & $-1.19$ & $-0.70$ & 19.94 & 0.85 & 2 & $-11.33$  &\cr
 378**& VCC 1286 & dE  & 12 30 24.61 & 12 47 35.2 &      & 20.57 & $-1.41$ & $-0.94$ & 19.92 & 0.99 & 1 & $-11.32$  &\cr
 379 &          & dE   & 12 25 37.54 & 17 50 36.9 &      & 20.62 & $-1.15$ & $-0.53$ & 19.94 & 0.77 & 2 & $-11.32$  &\cr
 380 &          & dI   & 12 46 17.91 & 11 11 09.4 &      & 20.91 & $-1.51$ & $-0.66$ & 19.99 & 0.90 & 1 & $-11.32$  &\cr
 381 &          & dI   & 12 35 07.06 & 11 39 37.6 &      & 21.61 & $-1.24$ & $-0.93$ & 20.00 & 1.00 & 2 & $-11.32$  &\cr
382* &          & dE/I & 12 22 39.27 & 18 05 20.1 &      & 20.98 & $-1.45$ & $-0.69$ & 20.01 & 0.90 & 2 & $-11.29$  &\cr
383  &          & dE,N & 12 18 14.47 & 16 44 08.2 &      & 21.57 & $-1.08$ & $-0.92$ & 19.99 & 0.95 & 1 & $-11.27$  &\cr
384  & VCC 1634 & dE   & 12 35 41.86 & 12 12 24.4 &      & 21.46 & $-0.94$ & $-0.84$ & 20.05 & 1.00 & 1 & $-11.27$  &\cr
385  &          & dE/I & 12 32 19.36 & 10 56 29.7 &      & 21.65 & $-1.11$ & $-0.93$ & 20.06 & 1.00 & 1 & $-11.27$  &\cr
386  &          & VLSB & 12 32 30.07 & 11 50 01.4 &      & 21.65 & $-1.29$ & $-0.94$ & 20.03 & 0.94 & 1 & $-11.26$  &\cr
 387**&         & dI   & 12 44 35.23 & 12 18 34.4 &      & 21.68 & $-1.24$ & $-1.44$ & 20.68 & 1.00 & 2 & $-11.26$  &\cr
 388 &          & dI   & 12 30 57.44 & 11 05 15.9 &      & 20.77 & $-1.31$ & $-0.55$ & 20.06 & 0.78 & 2 & $-11.23$  &\cr
 389 &          & dE/I & 12 24 51.16 & 15 23 39.4 &      & 20.93 & $-1.31$ & $-0.71$ & 20.04 & 0.88 & 1 & $-11.22$  &\cr
 390 &          & dE   & 12 44 39.14 & 12 19 14.7 &      & 20.66 & $-1.12$ & $-0.44$ & 20.06 & 0.76 & 2 & $-11.21$  &\cr
 391 &          & dE,N & 12 37 15.99 & 11 39 47.4 &      & 20.89 & $-1.01$ & $-0.59$ & 20.11 & 0.82 & 2 & $-11.20$  &\cr
 392 &          & dI   & 12 21 54.17 & 13 15 01.8 &      & 20.66 & $-1.08$ & $-0.40$ & 20.19 & 0.81 & 2 & $-11.20$  &\cr
 393** &        & dI   & 12 34 14.76 & 12 42 55.2 &      & 21.45 & $-1.45$ & $-1.64$ & 20.09 & 1.00 & 1 & $-11.20$  &\cr
 394**&         & dE/I & 12 40 01.63 & 11 52 45.3 &      & 20.97 & $-0.71$ & $-0.84$ & 20.09 & 0.93 & 1 & $-11.20$  &\cr
 395** &        & dE,N & 12 28 46.97 & 12 38 31.7 &      & 21.39 & $-0.82$ & $-1.05$ & 20.05 & 0.98 & 2 & $-11.19$  &\cr
 396 & VCC 1635 & dE   & 12 35 40.82 & 12 14 07.6 &      & 20.89 & $-1.17$ & $-0.58$ & 20.13 & 0.84 & 2 & $-11.19$  &\cr
 397 &          & VLSB & 12 22 22.52 & 14 25 50.3 &      & 21.81 & $-1.37$ & $-0.94$ & 20.18 & 1.00 & 1 & $-11.13$  &\cr
 398** &        & dI   & 12 30 28.27 & 12 58 57.6 &      & 20.80 & $-1.24$ & $-1.91$ & 20.14 & 1.00 & 2 & $-11.10$  &\cr
 399 &          & dE/I & 12 49 04.28 & 11 10 37.2 &      & 21.66 & $-1.18$ & $-0.88$ & 20.21 & 0.92 & 2 & $-11.07$  &\cr
\noalign{\smallskip \hrule}
\noalign{\smallskip}\cr}}$$}
\end{table*}

\begin{table*}
{\vskip 0.55mm}
{$$\vbox{
\halign {\hfil #\hfil && \quad \hfil #\hfil \cr
\noalign{\hrule \medskip}
(1)  & (2) &  (3) & (4) & (5) & (6) &
(7)  & (8) & (9) & (10) &
(11) & (12) & (13) &\cr
ID  & Name &  Type & $\alpha$ (J2000) & $\delta$ (J2000) & $V_h$/km s$^{-1}$ &
$B(6)$  & ICP & OCP & $B_T$ &
$P$ & Class & $M_B$ &\cr
\noalign{\smallskip \hrule \smallskip}
400**& VCC 1139 & dE   & 12 28 51.25 & 11 57 28.1 &      & 20.72 & $-1.26$ & $-1.39$ & 20.21 & 1.00 & 1 & $-11.05$  &\cr
 401 &          & VLSB & 12 20 24.62 & 16 04 22.2 &      & 21.26 & $-0.64$ & $-0.71$ & 20.24 & 0.80 & 1 & $-11.03$  &\cr
 402 &          & dI   & 12 41 14.05 & 12 14 58.8 &      & 21.33 & $-1.22$ & $-0.72$ & 20.28 & 0.81 & 1 & $-11.02$  &\cr
 403** &        & VLSB & 12 42 53.94 & 12 32 16.5 &      & 21.70 & $-1.21$ & $-1.03$ & 20.28 & 1.00 & 1 & $-11.00$  &\cr
 404 &          & dI   & 12 40 37.18 & 11 07 26.6 &      & 20.77 & $-1.16$ & $-0.42$ & 20.26 & 0.83 & 2 & $-11.00$  &\cr
 405** &          & dE/I & 12 44 24.22 & 12 10 25.4 &      & 21.48 & $-0.72$ & $-0.62$ & 20.27 & 0.83 & 2 & $-10.99$  &\cr
 406 & VCC 1631 & dE   & 12 35 38.34 & 12 20 29.9 &      & 20.88 & $-1.26$ & $-0.42$ & 20.33 & 0.85 & 1 & $-10.99$  &\cr
 407 &          & dE   & 12 39 57.47 & 12 06 47.4 &      & 21.13 & $-1.28$ & $-0.57$ & 20.38 & 0.82 & 1 & $-10.98$  &\cr
 408**&         & dI   & 12 37 19.38 & 11 52 12.5 &      & 21.22 & $-1.24$ & $-1.71$ & 20.37 & 1.00 & 1 & $-10.98$  &\cr
 409 &          & dE   & 12 33 45.32 & 10 52 19.7 &      & 21.17 & $-1.18$ & $-0.63$ & 20.31 & 0.88 & 2 & $-10.96$  &\cr
 410 &          & dE/I & 12 47 27.97 & 12 11 50.7 &      & 20.99 & $-1.25$ & $-0.52$ & 20.32 & 0.84 & 2 & $-10.94$  &\cr
 411 &          & dE/I & 12 30 53.75 & 10 54 43.0 &      & 21.48 & $-1.25$ & $-0.76$ & 20.36 & 0.87 & 1 & $-10.93$  &\cr
412**&          & dE/I & 12 37 03.28 & 11 25 09.9 &      & 21.66 & $-1.54$ & $-3.34$ & 20.37 & 1.00 & 1 & $-10.93$  &\cr
413**&          & dI   & 12 31 28.20 & 12 51 21.4 &      & 21.10 & $-1.35$ & $-1.49$ & 20.33 & 1.00 & 1 & $-10.92$  &\cr
 414*&          & dE   & 12 24 27.85 & 18 27 24.5 &      & 21.39 & $-1.17$ & $-0.69$ & 20.41 & 0.85 & 2 & $-10.90$  &\cr
 415*&          & dE/I & 12 45 10.07 & 11 41 20.2 &      & 21.04 & $-1.02$ & $-0.56$ & 20.36 & 0.80 & 2 & $-10.90$  &\cr
 416 &          & dE/I & 12 23 06.05 & 17 05 25.1 &      & 21.33 & $-1.14$ & $-0.67$ & 20.40 & 0.85 & 2 & $-10.86$  &\cr
 417 &          & dI   & 12 40 22.80 & 11 13 36.9 &      & 21.50 & $-1.05$ & $-0.74$ & 20.42 & 0.86 & 2 & $-10.86$  &\cr
 418 &          & dI   & 12 22 43.14 & 14 18 08.9 &      & 21.53 & $-1.14$ & $-0.71$ & 20.48 & 0.78 & 2 & $-10.85$  &\cr
 419** &          & dE   & 12 30 24.48 & 12 47 34.5 &      & 20.60 & $-1.44$ & $-0.62$ & 20.40 & 0.83 & 1 & $-10.84$  &\cr
 420 &          & dE/I & 12 18 34.86 & 18 35 49.3 &      & 21.15 & $-1.29$ & $-0.62$ & 20.45 & 0.82 & 2 & $-10.84$  &\cr
 421 &          & dI   & 12 34 36.14 & 11 04 23.1 &      & 21.20 & $-1.13$ & $-0.56$ & 20.47 & 0.82 & 2 & $-10.80$  &\cr
 422 &          & VLSB & 12 33 52.45 & 12 07 01.6 &      & 21.85 & $-1.24$ & $-0.84$ & 20.50 & 0.99 & 1 & $-10.80$  &\cr
 423 &          & dE/I & 12 49 23.81 & 11 12 06.6 &      & 21.25 & $-1.22$ & $-0.54$ & 20.50 & 0.79 & 2 & $-10.78$  &\cr
 424 &          & dI   & 12 43 21.77 & 11 00 18.8 &      & 21.55 & $-1.18$ & $-0.72$ & 20.52 & 0.75 & 2 & $-10.75$  &\cr
 425 &          & dE/I & 12 32 25.52 & 12 08 54.7 &      & 21.45 & $-1.20$ & $-0.65$ & 20.56 & 0.83 & 2 & $-10.72$  &\cr
 426 &          & dI   & 12 32 33.52 & 12 47 22.8 &      & 21.55 & $-1.20$ & $-0.51$ & 20.57 & 0.82 & 2 & $-10.69$  &\cr
 427 &          & dE/I & 12 38 06.47 & 12 17 51.8 &      & 21.44 & $-1.16$ & $-0.50$ & 20.66 & 0.85 & 2 & $-10.68$  &\cr
 428** &        & VLSB & 12 33 51.84 & 12 41 49.1 &      & 21.44 & $-1.54$ & $-1.21$ & 20.65 & 1.00 & 1 & $-10.63$  &\cr
 429** &        & dE/I & 12 42 45.07 & 11 22 43.1 &      & 20.29 & $-0.25$ & $-0.31$ & 20.64 & 0.00 & 2 & $-10.62$  &\cr
 430 &          & VLSB & 12 25 55.25 & 18 20 16.9 &      & 21.91 & $-1.45$ & $-0.80$ & 20.69 & 0.98 & 1 & $-10.61$  &\cr
 431 &          & dE   & 12 35 20.57 & 11 06 43.2 &      & 21.28 & $-1.26$ & $-0.44$ & 20.75 & 0.85 & 2 & $-10.51$  &\cr
 432 &          & dE/I & 12 38 00.41 & 11 34 29.6 &      & 21.60 & $-1.13$ & $-0.60$ & 20.80 & 0.83 & 2 & $-10.51$  &\cr
 433 &          & dE/I & 12 20 33.16 & 16 43 54.5 &      & 21.66 & $-1.13$ & $-0.64$ & 20.79 & 0.82 & 2 & $-10.48$  &\cr
 434 &          & dI   & 12 35 38.18 & 12 40 53.5 &      & 21.57 & $-1.24$ & $-0.52$ & 20.92 & 0.84 & 2 & $-10.43$  &\cr
 435 &          & VLSB & 12 32 37.82 & 11 24 44.1 &      & 22.41 & $-1.19$ & $-0.87$ & 20.98 & 1.00 & 1 & $-10.41$  &\cr
 436 &          & dE/I & 12 29 09.32 & 12 29 43.6 &      & 21.50 & $-1.34$ & $-0.48$ & 20.90 & 0.86 & 2 & $-10.35$  &\cr
 437 &          & dI   & 12 46 46.51 & 11 39 19.5 &      & 21.39 & $-1.36$ & $-0.41$ & 20.91 & 0.89 & 1 & $-10.35$  &\cr
 438 &          & dI   & 12 33 11.99 & 11 12 53.6 &      & 21.59 & $-1.10$ & $-0.45$ & 21.04 & 0.85 & 2 & $-10.35$  &\cr
 439 &          & dI   & 12 19 34.95 & 17 14 24.4 &      & 21.51 & $-1.12$ & $-0.44$ & 20.96 & 0.86 & 2 & $-10.31$  &\cr
 440 &          & dE/I & 12 30 01.86 & 12 56 52.8 &      & 21.40 & $-1.03$ & $-0.39$ & 20.93 & 0.88 & 2 & $-10.31$  &\cr
 441 &          & dI   & 12 29 01.10 & 12 33 32.8 &      & 21.69 & $-1.16$ & $-0.54$ & 21.01 & 0.84 & 2 & $ -9.67$  &\cr
 442 &          & dI   & 12 38 26.36 & 11 39 11.2 &      & 21.98 & $-1.12$ & $-0.62$ & 21.15 & 0.91 & 2 & $ -9.59$  &\cr
 443 &          & dE/I & 12 33 24.74 & 12 24 11.3 &      & 21.66 & $-1.07$ & $-0.42$ & 21.16 & 0.90 & 2 & $ -9.56$  &\cr
 444   &        & dI   & 12 38 47.03 & 12 14 17.0 &      & 21.97 & $-0.89$ & $-0.54$ & 21.28 & 1.00 & 2 & $ -9.52$  &\cr
 445   &        & dE/I & 12 35 29.48 & 12 40 59.6 &      & 21.84 & $-1.12$ & $-0.47$ & 21.27 & 0.86 & 2 & $ -9.51$  &\cr
 446** &          & dE/I & 12 42 12.67 & 12 22 16.5 &      & 21.71 & $-1.04$ & $-1.00$ & 21.36 & 1.00 & 2 & $ -9.38$  &\cr
 447   &        & dI   & 12 51 17.12 & 12 10 14.5 &      & 22.06 & $-1.33$ & $-0.52$ & 21.40 & 1.00 & 2 & $ -9.33$  &\cr
 448 &          & dE   & 12 32 47.48 & 11 18 06.8 &      & 21.97 & $-1.25$ & $-0.28$ & 21.66 & 1.00 & 2 & $ -9.17$  &\cr
 449 &          & dE/I & 12 49 30.24 & 11 13 05.9 &      & 22.34 & $-0.98$ & $-0.58$ & 21.80 & 0.98 & 2 & $ -8.93$  &\cr
\noalign{\smallskip \hrule}
\noalign{\smallskip}\cr}}$$}
\end{table*}

\section{The Luminosity Function}

\begin{table*}
\caption{The Virgo Cluster luminosity function}
{\vskip 0.75mm}
{$$\vbox{
\halign {\hfil #\hfil && \quad \hfil #\hfil \cr
\noalign{\hrule \medskip}
$M_B$ range  & Number & & $\log_{10} ({\rm N_{\rm gal}}$ mag$^{-1}$
deg$^{-2}$) & & $\alpha$ & &\cr
      &  0+1+2 & 0+1 & 0+1+2 & 0+1 & 0+1+2 & 0+1 &\cr
\noalign{\smallskip \hrule \smallskip}
\cr
$-22 < M_B < -21$  &  6 &  6 & $-0.62 \pm 0.18$ &  $-0.62 \pm 0.18$ & & &\cr
$-21 < M_B < -20$  &  4 &  4 & $-0.79 \pm 0.22$ &  $-0.79 \pm 0.22$ & $-1.32 \pm
 0.28$ &  $-1.32 \pm 0.28$ &\cr
$-20 < M_B < -19$  & 10 & 10 & $-0.40 \pm 0.14$ &  $-0.40 \pm 0.14$ & $-1.65 \pm
 0.28$ &  $-1.65 \pm 0.28$ &\cr
$-19 < M_B < -18$  & 15 & 15 & $-0.22 \pm 0.11$ &  $-0.22 \pm 0.11$ & $-0.95 \pm
 0.25$ &  $-0.95 \pm 0.25$ &\cr
$-18 < M_B < -17$  &  9 &  9 & $-0.44 \pm 0.14$ &  $-0.44 \pm 0.14$ & $-1.15 \pm
 0.19$ &  $-1.15 \pm 0.19$ &\cr
$-17 < M_B < -16$  & 19 & 19 & $-0.12 \pm 0.10$ &  $-0.12 \pm 0.10$ & $-1.64 \pm
 0.19$ &  $-1.60 \pm 0.20$ &\cr
$-16 < M_B < -15$  & 31 & 29 & $0.10 \pm 0.08$ &  $0.07 \pm 0.08$ & $-1.58 \pm 0
.14$ &  $-1.37 \pm 0.15$ &\cr
$-15 < M_B < -14$  & 55 & 38 & $0.34 \pm 0.06$ &  $0.18 \pm 0.07$ & $-1.49 \pm 0
.11$ &  $-1.37 \pm 0.12$ &\cr
$-14 < M_B < -13$  & 79 & 57 & $0.50 \pm 0.05$ &  $0.36 \pm 0.06$ & $-1.26 \pm 0
.09$ &  $-1.27 \pm 0.11$ &\cr
$-13 < M_B < -12$  & 90 & 64 & $0.56 \pm 0.05$ &  $0.41 \pm 0.05$ & $-1.02 \pm 0
.09$ &  $-0.90 \pm 0.11$ &\cr
$-12 < M_B < -11$  & 83 & 47 & $0.52 \pm 0.05$ &  $0.28 \pm 0.06$ &  &  &\cr
\noalign{\smallskip \hrule}
\noalign{\smallskip}\cr}}$$}
\begin{list}{}{}
\item[]The values of $\alpha$ in this table come from
power-law fits over a three magnitude range centered
on the middle of the magnitude range for the
appropriate entry.
\end{list}
\end{table*}

\begin{figure}
\begin{center}
\vskip-4mm
\psfig{file=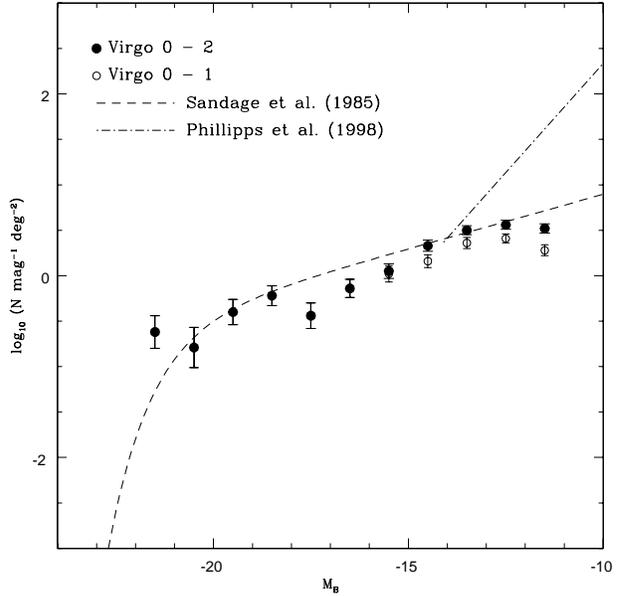, width=8.65cm}
\end{center}
\vskip-3mm
\caption{
The luminosity function for the Virgo Cluster sample. 
Filled circles represent the luminosity function for all galaxies
rated 0--2.  Open circles represent the luminosity function for all galaxies
rated 0--1.  The dashed line represents the Schechter (1976) function
fit of Sandage et al.~(1985) for all galaxies with $B < 20$ 
($M_B < -11.1$), including
their incompleteness corrections.
The dotted-dashed  line represents the power-law 
fit of Phillipps et al.~(1998a) to their
inner-area sample for all galaxies with $15.5 < R < 20$
($-15.6 < M_R < -11.1$),
scaled horizontally assuming $B-R=1.5$. 
The two fits are scaled vertically to have the same number of galaxies
as our current sample at $M_B = -14$. 
}
\end{figure}

The luminosity function for our galaxy sample is presented in
Figure 5 and Table 3.  

Our luminosity function is similar to that derived by Sandage et al.~(1985).
This is not surprising given  
that most of our galaxies were in the Virgo Cluster Catalog (except at
the faint end where our survey was deeper or the VCC is incomplete) and
the fact that our total magnitudes for the VCC galaxies (Fig.~4) are
similar to the VCC ones.

However, the difference with the results of Phillipps et al.~(1998a) is
very marked.  Our survey reached a similar depth to theirs, but in
the faintest three magnitude bins (Fig.~5) we found far fewer galaxies
per bright galaxy.  In our faintest bin we found only 83 galaxies, whereas
given the Phillipps et al.~(1998a) inner area LF normalized to our data
at $M_R = -14$ (the bright end of the magnitude range over which they fit
the LF), we would predict about 1000 galaxies.  Additionally, we found
that our LF steepened at brighter magnitudes: $M_B = -16.5$ compared with
$M_B = -13.5$ ($M_R = -15$) for the Phillipps et al.~(1998a) inner area LF.  
This discrepancy can be explained if the Phillipps et al.~(1998a) 
sample suffers from severe background contamination at the faint end.
Alternatively it could means that our sample is incomplete (it would
need to be incomplete at 
about the 90\% level) at the faint end relative to the
Phillipps et al.~(1998a) sample.  Such incompleteness could 
follow from us (and Sandage et al.~1985)
erroneously rejecting the higher surface-brightness
cluster members from our sample
because we think that they are background galaxies.
We will return to this point in Section 7. 
Finally, we note that the final two points in the
Phillipps et al.~(1998a) LF may have been overestimated by a factor of two,
meaning that a faint-end slope of $\alpha \sim -1.9$ may be appropriate
for that dataset, not $\alpha \sim -2.2$ (S.~Phillipps, private
communication).  This would make the inconsistency with the current
dataset smaller. 
 
Table 3 also gives $\alpha$, the logarithmic slope of the luminosity
function, at each absolute magnitude.
The curvature in the LF is real and statistically significant:
a single value of $\alpha$, independent of absolute magnitude,  
is highly inconsistent with the data,
Neither a power-law or a Schechter (1976) 
function provides a satisfactory fit to the data over any appreciable
magnitude range.
That we are able to make such a statement follows from the small (Poisson)
error bars in Table 3, which in turn follows from the large number of
galaxies in our sample.   
It is interesting, however, that the average value of $\alpha$ fainter than
$M_B = -18$ is $\alpha = -1.35$, close to the faint-end slope that
Sandage et al.~(1985) found for the VCC.
Over the magnitude range $-17 < M_B < -14$ the average slope is more like
$\alpha = -1.7$, similar to the slope in the Phillipps et al.~(1998a) sample
over this magnitude range, allowing for a $B-R$ colour of 1.5.  

Over the half-magnitude interval $-11 < M_B < -10.5$, 
only have 29 galaxies classified 0+1+2,
implying that the logarithm of the LF at $M_B = -10.75$ is
$0.37 \pm 0.08$  mag$^{-1}$ deg$^{-2}$. 
This could be a sign of a weak turnover in the LF, but there are other
ways to explain the paucity of galaxies in this interval.
For example, the sample could be incomplete at these faint levels (only
a modest amount of incompleteness would be required to generate this kind of
feature in the LF.  This kind of incompleteness could follow from the
reduced dynamic range in surface brightness over which we classify galaxies
1 or 2 at these faint levels (Trentham \& Tully 2001).  Galaxies with
surface-brightnesses a little brighter than about 27 $B$ mag arcsec$^{-2}$
will be missing from the sample because they are indistinguishable from
field galaxies of the type seen in large number in our blank fields.
Galaxies with
surface-brightnesses a little fainter than about 27 $B$ mag arcsec$^{-2}$
will also be missing from the sample because they are not detected
above the sky.  Consequently at these very faint levels our sample
may be incomplete. 
 
\section{Properties of galaxies}

\begin{figure}
\begin{center}
\vskip-4mm
\psfig{file=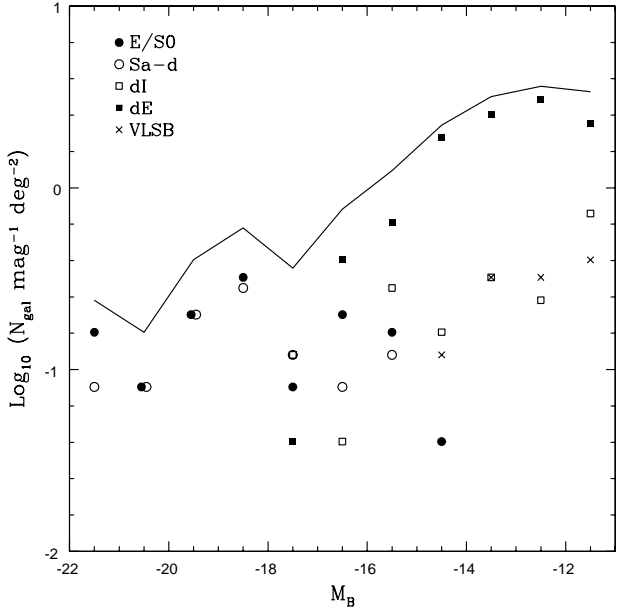, width=8.65cm}
\end{center}
\vskip-3mm
\caption{
The luminosity function segregated
by morphological type (galaxies classified dE/I or dS0 in Table 2 are
grouped with the dEs here and galaxies classified as BCD in Table 2 are
grouped with the dIs here).  
The line indicates the total luminosity function, as presented with
uncertainties in Figure 5.
}
\end{figure}

\begin{figure}
\begin{center}
\vskip-4mm
\psfig{file=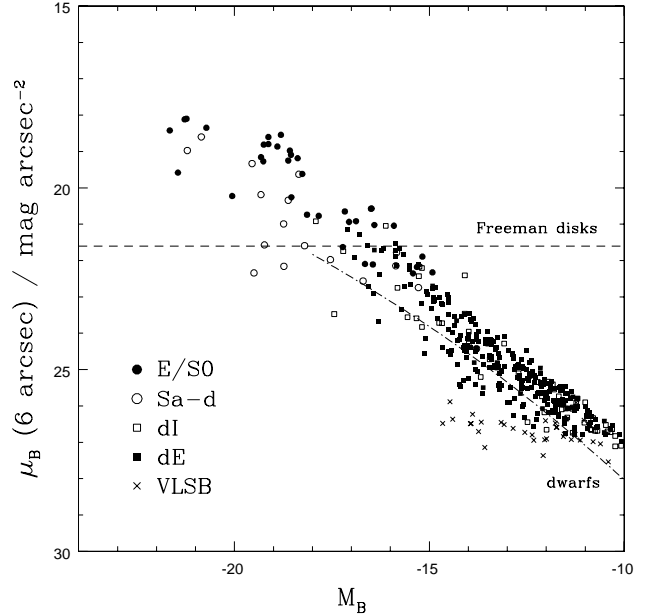, width=8.65cm}
\end{center}
\vskip-3mm
\caption{
Surface-brightness $\mu_B$ vs.~absolute magnitude $M_B$
for the sample galaxies, segregated
by morphological type (galaxies classified dE/I or dS0 in Table 2 are
indicated dE here and galaxies classified as BCD in Table 2 are
indicated dI here).  Surface-brightnesses 
are the average values measured within circular apertures of radius
6 arcseconds.  Galaxies with 
companion stars or other galaxies projected within
this aperture are not included.
The value of $\mu_B$ = 21.6 arc sec$^{-2}$ for galaxy disks (Freeman 1970)
is represented by the dashed lines; many spiral galaxies are above this
line due to the presence of a bulge.  The dotted-dashed line represents
typical values (e.g.~Binggeli \& Cameron 1991, Binggeli 1994) for 
exponential dwarf
galaxies; many dE galaxies are above this
line due to the presence of a nucleus and many dI galaxies are above
this line due to the presence of star-forming knots within the 
aperture. 
Both dwarf ellipticals
and dwarf irregulars have similar scaling laws
(Binggeli \& Cameron 1991)}
\end{figure}

\begin{figure}
\begin{center}
\vskip-4mm
\psfig{file=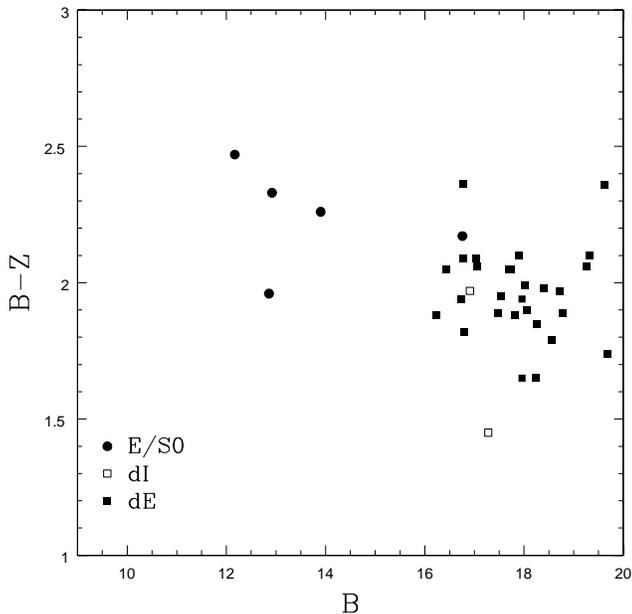, width=8.65cm}
\end{center}
\vskip-3mm
\caption{
Color-magnitude diagram for the 35 galaxies that were in the 9 fields
observed in $Z$ and bright enough to be 
detected at the 5$\sigma$ level in the $Z$
data.  All magnitudes were measured in an aperture of radius 6 arcseconds,  
which was large enough that the effect on the colours 
caused by differential seeing 
between the $B$ and $Z$ data was negligible.  
Only galaxies with total absolute magnitudes typically $M_B < -12$ 
are shown.  Fainter (and consequently lower surface brightness galaxies)
were not detected above sky in the $Z$-band data, particularly since
that data suffers from serious fringing, typically 6 per cent of sky
(http://www.ast.cam.ac.uk/~wfcsur/defringing.html).
Dwarf elliptical galaxies have
$B-Z$ colours of 1.7 -- 2.2
given the spectral
energy distributions of Trentham et al.~1998b)}
\end{figure}

The contribution to the total galaxy LF 
from galaxies of different morphological types
is presented in Figure 6. 

At the faint end, the vast majority of galaxies in the sample are dE
galaxies, as identified on morphological grounds.  The structural
parameters (see Figure 7) 
and  colours (see Figure 8) 
of the galaxies are consistent with
this interpretation. 
Many of these dE galaxies are nucleated, but few of the dIrr galaxies
in the cluster are.  This suggests either that the two kinds of galaxies
form in different ways (despite the similarity in their scaling laws) or
that dIs are dEs in formation and that the nucleus is the last part to
form.  
In the faintest three bins about one-third of the galaxies are rated 2.
These tended to be high surface-brightnesses dEs that could conceivably
be background late-type galaxies.  Even excluding these galaxies, dEs
are still the dominant types at the faintest magnitudes.    

The VLSB galaxies 
never contribute significantly to the total luminosity function.
This result cannot be directly compared to the 
measurements of VLSB galaxies in Fornax of Kambas et al.~(2000) because 
we require a somewhat
lower surface brightness (see Section 4)
than do Kambas et al.~(see Section 2.2 of their  
paper) to call a galaxy a ``VLSB''.  Nevertheless, we do find far fewer
low surface-brightness galaxies that are likely to be cluster members
than they did in Fornax.  
The Virgo Cluster has a crossing times shorter than 
one-tenth of a Hubble time (see
Table 1 in Tully et al.~1996) so galaxies here must have
undergone many galaxy-galaxy interactions.  
It is therefore surprising  
that diffuse VLSB galaxies can survive.   Perhaps this
is evidence that small galaxies in clusters have substantial dark matter
halos.

We also notice a paucity of galaxies having exactly $M_B = -17.5$.  A
dip in the LF at this absolute magnitude was also seen in the Coma Cluster
(Trentham 1998a) and in the NGC 1407 Group (Trentham \& Tully 2001),
another dense knot of early-type galaxies.  This is also the
magnitude where the cluster population changes from being dominated by
high surface-brightness giant elliptical galaxies to low surface-brightness
dwarf elliptical galaxies (see Figure 6).  These two types of stellar
system have very different structural parameters (see Figure 1 of Binggeli
1994), implying that they had very different formation mechanisms.

\section{Caveats: incompleteness and contamination}

We expect to be missing two sorts of galaxies from our sample: galaxies
with very low surface brightnesses that are never visible above the sky
and cluster galaxies with high surface brightness that we reject from the
sample because we think that they are background galaxies.  
We will argue that neither of these are likely to be a serious problem
with the current dataset.

Galaxies with very low surface brightnesses ($\mu_B > 27$ mag arcsec$^{-2}$
within an aperture of radius 6 arcseconds) are unlikely to be very common
for $M_B < -11$.  Evidence supporting this is
\vskip 1pt
\noindent
(i) such galaxies were not found in deep images of the Virgo Cluster
core (Trentham \& Tully 2001) taken with the 8 m Subaru Telescope.
Those images reached very deep surface brightness limits (about 28 $R$
mag arcsec$^{-2}$, equivalently $\sim$ 29 $B$ mag arcsec$^{-2}$  
for dwarf galaxies) but did not uncover any additional galaxies with
$M_B < -11$ that were missing from the current survey because their
surface brightnesses were too low;
\vskip 1pt
\noindent
(ii) In Figure 7, the points at the faint $M_B$ end do not cluster 
right at the very limit of detection (27 mag arcsec$^{-2}$).  Most
have higher surface brightnesses (although a few have 
$\mu_B \sim 27$ mag arcsec$^{-2}$).  This in turn suggests that there
do not exist large numbers of galaxies with surface brightnesses just
below this limit;  
\vskip 1pt
\noindent
(iii) No local galaxies with $M_B < -11$
are known with such low surface brightnesses, 
although a small number of bulge-dominated galaxies have disk components
this faint or fainter e.g.~Malin 1 (Bothun et al.~1987) 
and GP 1444 (Davies, Phillipps
\& Disney 1988).
The giant galaxy population in the Virgo Cluster is similar to that in
the local Universe in terms of galaxy structural parameters, so it
is reasonable to expect that the dwarf galaxy population is similar too
and that this absence in the field of dwarfs with $M_B < -11$ and
$\mu_B$ $>$ 27 mag arcsec$^{-2}$ extends to the Virgo Cluster.  
 
Dwarf galaxies with very high surface brightnesses lacking an extended
diffuse light component are another potential source of contamination
because they would be rejected from the current sample since they
look like luminous background galaxies (they have low $P$ values).
These can either be blue HII galaxies like Markarian 1460
(Trentham, Tully \& Verheijen 2001b) or red compact 
dwarfs like M32 in the Local Group or UGC 6805 in the Ursa Major Cluster
(Tully \& Verheijen 1997).  In the Virgo Cluster, VCC 1313 is an
example of the former and VCC 1627 of the latter; were the velocities
of these two objects not known we would have classified them as
background objects. 
For $M_B < -14$ only these two galaxies of this type were identified.
Were such objects to exist with $M_B > -14$,  
our sample could be incomplete.  
The fraction of detected galaxies that we
assign to the cluster on surface
brightnesss grounds
ranged from close to 1 at the bright end of the sample
($B < 15$) to about 1 per cent at the faint end ($B \sim 20$).

This is an important change of emphasis from what has classically been
thought of as the main uncertainty in studies of cluster luminosity
functions.  Previously 
the results from this kind of study were open to question
because many very
low surface-brightness galaxies could be missing 
from the sample 
since they are never detected above the sky.  This is no longer
a worry because deep surveys like this one and that
of Trentham \& Tully (2001) are not uncovering large numbers of
LSB galaxies that were missing in shallow ones.  Instead, the major concern
is now that the sample may be missing many {\it high} surface-brightness
galaxies which we have culled from the sample because we think
that they are background galaxies.

We do not however regard this as a serious worry.
Firstly, at the bright end of our sample, where velocity measurements
are available, galaxies with high surface brightnesses are rare (VCC 1313
and VCC 1627 are the exceptions).
Secondly, high surface-brightness galaxies
do not appear to be present in substantial number in
the Fornax Cluster else they would have been seen in the spectroscopic
survey described by Drinkwater et al.~(2000a; this work uses
the 2dF spectrograph on the Anglo-Australian Telescope).
Small numbers of compact galaxies 
with early-type spectra {\it were} discovered in that 
survey (Drinkwater et al.~2000b, Phillipps et al.~2001)
but they are too rare to contribute
significantly to the Fornax LF.  Given the similarities between the Virgo
and Fornax Clusters, we do not expect this to be a major source of
incompleteness in the current study.   
 
\begin{figure}
\begin{center}
\vskip-4mm
\psfig{file=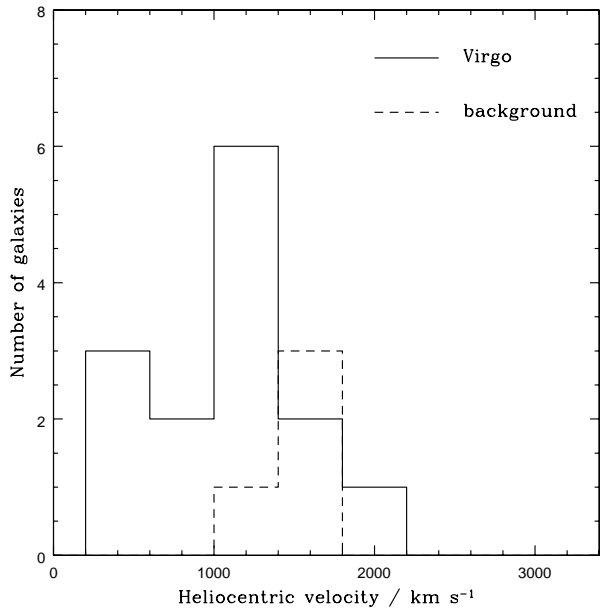, width=8.65cm}
\end{center}
\vskip-3mm
\caption{
Histogram of all galaxies in the Virgo and North Galactic Cap
background regions listed in the {\it Nearby Galaxies Catalog} (Tully 1987) and
having $M_B < -16$.  The peak in the Virgo sample at about 1000 km s$^{-1}$
represents the Virgo Cluster. 
}
\end{figure}

Another potential problem is that the Virgo data but not the North Galactic Cap
background data is contaminated by an anomalously large number of nearby
galaxies at comparable distances to the Virgo Cluster that are not bound to
the cluster
(the $P$ values computed by comparing the numbers of low
surface-brightness galaxies in the two datasets would then be too high).  
The contaminating galaxies would need to be nearby or else they would not
look like Virgo Cluster galaxies.
We regard such a possibility as unlikely due to the paucity of luminous
galaxies
with velocities between about 2000 km s$^{-1}$ and 3000 km s$^{-1}$ in the
Virgo dataset (see Figure 9).  Were such objects to be numerous, the
satellite populations of these luminous galaxies (only very luminous
galaxies are listed in the {\it Nearby Galaxies Catalog} at these
distances and so would be included in Figure 9) could look like Virgo Cluster
members (satellites of galaxies at higher velocities would be too small).

\section{The total optical luminosity in galaxies}

\begin{figure}
\begin{center}
\vskip-4mm
\psfig{file=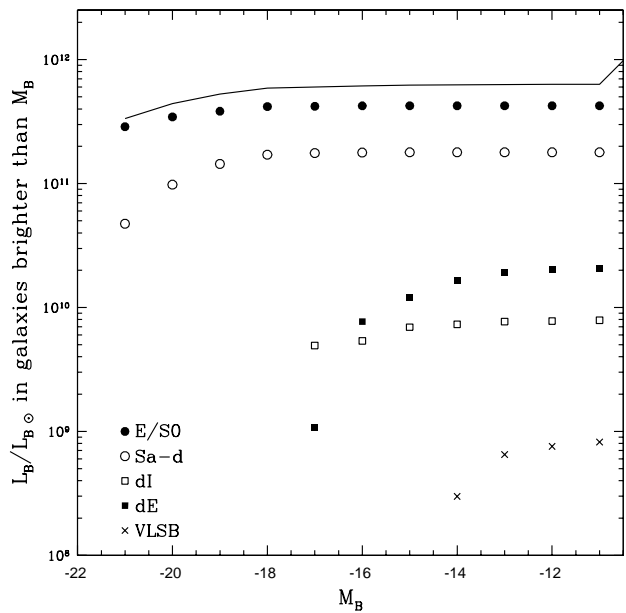, width=8.65cm}
\end{center}
\vskip-3mm
\caption{
The total blue luminosity in galaxies brighter than $M_B$, segregated
by morphological type (galaxies classified dE/I or dS0 in Table 2 are
indicated dE here and galaxies classified as BCD in Table 2 are
indicated dI here).
The line indicates the total luminosity in galaxies of all
morphological types.
}
\end{figure}

In Figure 10, we present the contribution from galaxies of different types
and absolute magnitudes to the total optical luminosity
of the region of the Virgo Cluster that we surveyed.  
The total luminosity in our sample is $6.3 \times 10^{11} 
{\rm L}_{\odot {\rm B}}$,
corresponding to a luminosity surface density of 
$5.6 \times 10^{10} {\rm L}_{\odot {\rm B}} {\rm Mpc}^{-2}$.   
For comparison, for the VCC, Sandage et al.~(1985, converted to the distance
scale used elsewhere in this paper) measure a
luminosity surface density of 
$3.4 \times 10^{10} {\rm L}_{\odot {\rm B}} {\rm Mpc}^{-2}$ averaged over
the central six degrees of the cluster. 
That our number is slightly higher follows from the fact that by
proportion our survey covers more high-density areas within the cluster
than does the VCC (see Figure 1).

Figure 10 shows that
only a small proportion (less than one-tenth) of the total optical
luminosity of the cluster is in dwarf galaxies.  Even less is in the
systems that we called VLSB galaxies.  
Were these galaxies not to be resolved as individual objects (as would be
the case if they were in a more distant cluster), they would be 
observable only though their  
contribution to the diffuse intracluster light.  
This in turn implies that if the cluster LF does not
vary strongly between clusters (see the next section), then the intracluster
light in distant clusters (typically $>$ 10 per cent of the total
cluster light e.g.~Melnick, Hoessel \& White 1977;
Thuan \& Kormendy 1977;
Scheick \& Kuhn 1994; V{\'{i}}lchez-G{\'{o}}mez, Pell{\'{o}} \&
Sanahuja 1994) cannot be produced by the integrated light from
low surface-brightness dwarf galaxies and is more likely to be
made up of stars
tidally released from luminous galaxies within the cluster, perhaps
via galaxy harassment (Moore et al.~1996).  

\section{Comparison with other LFs}

\begin{figure}
\begin{center}
\vskip-4mm
\psfig{file=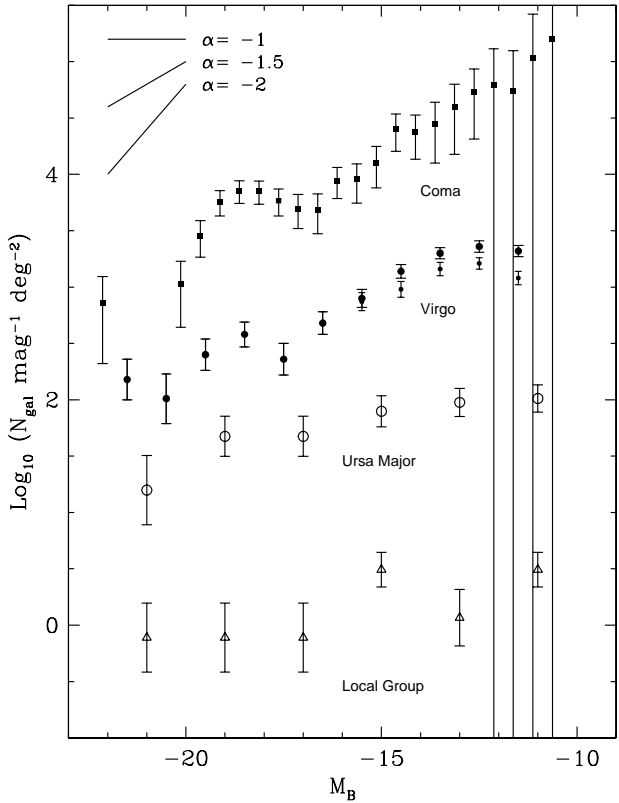, width=8.65cm}
\end{center}
\vskip-3mm
\caption{
The $B$-band luminosity functions of the Local Group
(open triangles), the Ursa Major
Cluster (open circles), the Virgo Cluster
(filled circles), and the Coma Cluster (filled squares).
The Virgo Cluster data are from this work, shifted
upward
in the ordinate axis by 2.8 units (to permit
clearer presentation).
The Ursa Major
data are from Trentham et al.~(2001a).  
The Coma Cluster data are from Trentham (1998a), shifted upward
in the ordinate axis by 1.5 units.
The Local Group data is from the compilation of Irwin
(http://www.ast.cam.ac.uk/$^{\sim}$mike/local$_{-}$members.html), 
adjusted to the
$B$-band (where independent photometric measurements do not
exist in the NED 
database) assuming $B-V=0.6$ for the Milky Way
(see the template Sbc galaxy spectral energy distribution
of Coleman, Wu \& Weedman 1980), $B-V=0.3$ for
dwarf irregular galaxies (Coleman et al.~1980), and $B-V=0.8$ for 
dE/dSph galaxies (Caldwell 1983).  This Local Group data is
shifted upward on the ordinate axis by 5.5 units.
}
\end{figure}

\begin{table}
\caption{Dwarf-to-giant ratios for various environments}
{\vskip 0.75mm}
{$$\vbox{
\halign {\hfil #\hfil && \quad \hfil #\hfil \cr
\noalign{\hrule \medskip}
Environment  & ${{{\rm N}(-16 < M_B < -11)}\over{{\rm N}(M_B < -16)}}$ &\cr 
\noalign{\smallskip \hrule \smallskip}
\cr
Coma  & $11.73 \pm 6.76$  &\cr
 & & & & &\cr
Virgo (0+1+2) & $5.36 \pm 0.74$ &\cr 
Virgo (0+1)   & $3.73 \pm 0.53$ &\cr  
 & & & & &\cr
Ursa Major   & $2.07 \pm 0.67$  &\cr
 & & & & &\cr
Local Group  & $2.83 \pm 1.35$ &\cr
\noalign{\smallskip \hrule}
\noalign{\smallskip}\cr}}$$}
\end{table}

In Figure 11 we compare the Virgo Cluster LF to the LFs for the Coma
Cluster ($z=0.023$), the Ursa Major Cluster, and the Local Group.  In Table 4
we present the dwarf-to-giant ratios (DGRs) 
for the various
environments.  The DGR is a convenient way to parameterize the LF
by a single number; note that different authors e.g.~Phillipps et 
al.~(1998b) define this quantity in a different way.
The Virgo LF presented here and the Ursa Major LF
(Trentham, Tully \& Verheijen 2001a) are somewhat more tightly
constrained than the Coma LF or the Local Group LF.
For the Coma Cluster, the error bars are large due to the need to determine
the LF using a background subtraction and the large field-to-field variance
of the background (the Coma Cluster is sufficiently distant that the faint
dwarfs become smaller than the seeing and we cannot establish membership on
the basis of morphology, as we did in the current study).   
For the Local Group, the error bars are large due to Poisson counting
statistics, since there are not many galaxies (there are only six
galaxies -- M31, the Milky Way, M33, IC 10, LMC and SMC --   
brighter than $M_B = -16$ for example). 

There appear to be two types of galaxy luminosity function -- one for
evolved regions (where the elliptical galaxy fraction is high, the
galaxy density is high and the crossing time is short) like the Virgo 
Cluster and Coma Cluster and one for unevolved regions like the
Ursa Major Cluster and the Local Group.  The Virgo and Ursa Major LFs
represent natural prototypes for the two kinds of LF.
The two LFs
are inconsistent with each other at a high level of significance: 
the probability that the two LFs are drawn from a single distribution
is $<< 1$ per cent (reduced $\chi^2 = 22.2$ for 5 degrees of freedom).
The reason that such a strong statement can be made follows from the
small error bars for these two LFs on Figure 11.

The evolved region LF appears to be characteristic of many clusters
(see Trentham 1998c), even though for each cluster individually (like Coma) the
LF is poorly determined due the large uncertainties following a background
subtraction.
In the study of Trentham (1998c), the composite cluster LF was determined
primarily from the LFs of Virgo (Sandage et al.~1985) and Fornax (Ferguson
\& Sandage 1988).  The current LF, at least in all but the
faintest one or two magnitude bins, can to some extent be seen as a
verification of the Virgo LF of Sandage et al.~(1985) and therefore
the LF presented in Table 3 may be regarded as being valid (when
scaled appropriately) for the majority of galaxy clusters.  
The main features of this evolved LF are the steep rise at $M_B=-16$
and the flattening faintward of $M_B=-14$.
In the very centres of the richest clusters, the galaxy density is high
enough that many dwarfs may be destroyed via cluster-related processes
(e.g.~Phillipps et al.~1998b, Adami et al.~2000, Boyce et al.~2001) and
the rise at $M_B=-16$ is no longer observable there, but the Virgo
Cluster is unlikely to be rich enough for this phenomenon to be
important. 
 
The unevolved (i.e.~Ursa 
Major) LF is different from the evolved one in that it lacks the
rise ($\alpha = -1.6$) at $M_B = -16$ and consequently generates
a lower DGR.  This LF ($\alpha \sim -1.1$ everywhere fainter
than $M_B =-18$) is also appropriate for the Local Group (van den
Bergh 1992, 2000) and for
the true field, where the LF is determined spectroscopically (see Table 5;
the values of $\alpha$ quoted by different authors are derived in
different ways, for example by Schechter (1976) function fits with
different $L^{*}$,  but none are as steep as the Virgo LF at around $M_B=-16$). 

\section{Comparison with theory}

\begin{table}
\caption{Faint-end slopes of spectroscopic field surveys}
{\vskip 0.75mm}
{$$\vbox{
\halign {\hfil #\hfil && \quad \hfil #\hfil \cr
\noalign{\hrule \medskip}
Survey & $\alpha$ & Reference&\cr
\noalign{\smallskip \hrule \smallskip}
\cr
Stromolo-APM  & $-1.0$ & Loveday et al.~1992  &\cr
Hawaii-Caltech               & $\sim -1.25$ & Cowie et al.~1996    &\cr
Autofib                      & $-1.1$ & Ellis et al.~1996    &\cr
LCRS & $-0.7$ & Lin et al.~1996      &\cr
Sloan                        & $-1.2$ & Blanton et al.~2001    &\cr
2dF + 2MASS (Near-IR) & $-1.0$ & Cole et al.~2001    &\cr
\noalign{\smallskip \hrule}
\noalign{\smallskip}\cr}}$$}
\end{table}

The discussion in the previous section can be summarized by the
following results:
\vskip 1pt \noindent
1) The Virgo Cluster LF has a rise with $\alpha = -1.6$ at
$M_B = -16$ than is not seen
in the LFs of the field  
or unevolved environments like the Ursa Major Cluster;
\vskip 1pt \noindent
2) This rise does not continue indefinitely towards fainter magnitudes
and the Virgo Cluster LF is flat ($\alpha \sim -1.0$) between $M_B = -14$
and the limit of our survey at $M_B = -11$.  
\vskip 1pt \noindent
Furthermore,
\vskip 1pt \noindent
3) Dwarf galaxies are very deficient in the Virgo Cluster compared to
the predictions of CDM theory (which predicts $\alpha \sim -2$ if light
traces mass).  The implication of 1) is then that dwarfs are even more
deficient in the field. 

Attempting to explain these three results in the context of galaxy
formation models leads us to consider the following four physical processes:
\vskip 1pt \noindent
(i) Tully (2001; see also Somerville 2001 and Tully et al.~2001) has
suggested that the dark halos in dense, evolved environments formed early
in the history of the Universe, prior to reionization, but that the
dark halos in diffuse, unevolved environments assembled much later.
The dark halos in the evolved environments, like the Virgo Cluster,
could then collect gas which could later be turned into stars, but 
dark halos in unevolved environments could not -- in the terminology
of Tully (2001), the formation of stars within these halos would 
be ``squelched''.  
The end result would be more dwarf galaxies per unit total mass
in evolved environments than in unevolved environments, in agreement
with result 1) above.  In unevolved environments there would be very
many dark matter halos with no stars at all.  Such a phenomenon is
also seen in simulations 
(Chiu, Gnedin \& Ostriker 2001) ;  
\vskip 1pt \noindent
(ii) Local feedback can generate a very low star formation efficiency
in low mass galaxies.  Winds from a
modest number of supernovae can expel a large
fraction of the gas in small galaxies (Dekel \& Silk 1986,
Efstathiou 2000) because the
galaxies have small potential wells.  The consequence of this is that
the final luminosity $L$ of a small galaxies is very strongly decreasing
function of the galaxy mass $M$ so that $M/L$ is a decreasing function
of $M$.  The LF that we measure is then much shallower than the mass
function, which is what CDM predicts.  That this process is important is
suggested by the observation that the lowest mass dwarf galaxies   
are heavily dark matter dominated (Wilkinson et al.~2001). 
This mechanism 
solves the discrepancy with CDM theory in a different way from squelching
-- it causes all low-mass galaxies to have very low luminosities, rather
than turning off star formation altogether in all but very few low-mass
halos, leaving most low-mass halos completely dark; 
\vskip 1pt \noindent
(iii) Cluster-related processes like galaxy harassment (Moore et al.~1996)
and tidal interactions during the early stages of cluster
formation can form dwarfs (Barnes \& Hernquist 1992)
and may in part be responsible for the higher
DGR in clusters.
On the other hand, dwarfs formed this way are not be expected to have
dark halos (and by implication, low surface brightnesses) like the
Virgo dwarfs we studied here; the dwarfs found in the simulations of
Barnes \& Hernquist do not have appreciable dark-matter content.  Additionally,
field and cluster dwarfs seem to
have similar scaling laws (see Figure 7), which would seem to argue that
field and cluster dwarfs formed in a similar way to
each other so that cluster-related
processes are probably not the main mechanisms 
responsible for the difference in the
evolved and unevolved LFs;  
\vskip 1pt \noindent
(iv) the number of dwarfs predicted by theory is very much lowered if
the power spectrum $P(k)$ of primordial fluctuations is reduced on small
scales.  A general phenomenon of CDM theory is its success on large scales
but its failure to reproduce observations on small ($<10$ kpc) scales --
the two most serious failures are the dwarf galaxy deficiency studied
here and the observed flatness of dark matter profiles in the centres of
galaxies (e.g.~Binney, Gerhard \& Silk 2001).  Possible mechanisms to 
remove small-scale power from the CDM fluctuation spectrum include
making the dark matter warm (Bode, Ostriker \& Turok 2001) or 
self-interacting (Spergel \& Steinhardt 2000, however see
Miralda-Escud{\'{e}} 2001 and Moore et al.~2000).  
Studying the effects of these particular modifications to the dark-matter
power spectrum is currently an active field of study.

In Table 6 we summarize how the different physical processes can explain
the observational results listed above.  A check is placed in a column
whenever an observational result follows from the inclusion of the physical
process in question in models of galaxy formation.  

\begin{table*}
\caption{Confrontation of observation with theory}
{\vskip 0.75mm}
{$$\vbox{
\halign {\hfil #\hfil && \quad \hfil #\hfil \cr
\noalign{\hrule \medskip}
Process & Steepening of Virgo & Flattening of Virgo & Deficiency of dwarf &\cr  
& but not UMa/field   & LF faintward of $M_B=-14$ & galaxies compared to &\cr 
 &          LF at $M_B=-16$ & & CDM predictions &\cr  
\noalign{\smallskip \hrule \smallskip}
\cr
 & & & &\cr
Squelching & $\surd$ &  & $\surd$ &\cr  
 & & & &\cr 
Supernova feedback & & $\surd$ &  $\surd$ &\cr   
 & & & &\cr 
Cluster-related processes & $\surd$ &  &  &\cr   
 & & & &\cr 
Reducing $P(k)$ on small scales & & & $\surd$ &\cr   
 & & & &\cr
\noalign{\smallskip \hrule}
\noalign{\smallskip}\cr}}$$}
\end{table*}

\section{Future work}

A natural extension of the current study is 
the determination of a deeper and more
accurate Virgo Cluster LF.
For $M_B < -11$, the main source of uncertainty in the current work
is in establishing
membership, not in counting statistics: the difference between the 0--1 and
0--2 points in Figure 5 is greater than the error bars.
Galaxies classified ``2'', the ones for
which we are uncertain about membership, tend
to be the higher surface-brightness ones in our sample,
so it should be possible to
establish distances to these either by measuring spectroscopic  
redshifts or from surface-brightness 
fluctuations (Jerjen, Freeman \& Binggeli 1998,
2000). 
For $M_B > -11$, deeper observations than those presented here over
similarly large angular areas will be
required.
Such observations are now possible with the advent of mosaic CCDs on
8 m telescopes like Suprime-cam on the Subaru Telescope (for deep observations
of 1.2 deg$^{2}$ in the core of the Virgo Cluster, see Trentham \& Tully 2001). 
 
A spectroscopic survey of the Virgo Cluster could be extended to
include significant numbers of compact, high surface-brightness galaxies
that we think are background, in the style of the Drinkwater
et al.~(2000a) Fornax survey.  If the vast majority of such objects are
indeed background objects, this would alleviate the concern highlighted in
Section 7 that the current sample is heavily incomplete due to us  
rejecting such galaxies.  Such a project is now feasible with
the advent of wide-field multi-object spectrographs on large telescopes.

More detailed studies of the dwarf galaxies found in Virgo will also be of
value in assessing the importance of the various physical processes at
work during galaxy and cluster formations.  The following observations should
be of particular use:
\vskip 1pt \noindent
1) measurement of colours and elemental abundances.  These will constrain the
star-formation histories of the 
Virgo dwarfs which in turn will allow a lower limit to be placed on the
redshift at which gas was collected by small dark-matter halos.  This is of
importance in the context of the squelching
picture described above.  This squelching
picture also relies on the very existence of
these dark-matter halos around the Virgo dwarfs.
In the long term, studies of the kinematics
of stars in the dwarfs will be required to verify this assertion;
\vskip 1pt \noindent
2) HI measurements of the cold gas content of the Virgo dwarfs.  Galaxies
that have stayed any appreciable time in the Virgo Cluster would lose
their gas via ram-pressure stripping from the
cluster X-ray halo.  Therefore a large number of dwarfs
in Virgo with HI would suggest that many Virgo dwarfs only recently
entered the cluster.  This would in turn argue against any physical
process that requires Virgo dwarfs to have formed in the cluster or
at very early times in small groups which quickly merged to form the 
cluster;
\vskip 1pt \noindent
3) the location of dwarfs in the cluster.  Do giants maintain their 
dwarf populations once they are in the Virgo Cluster, or do the dwarfs
adopt orbits determined by the cluster potential?  If the former were
true, dwarfs would tend to cluster around giants.  If the latter were
true, they would be smoothly distributed throughout the cluster, with
a radial density profile similar to that of the giant galaxies.  
The answer to this question will provide constraints on the dark-matter
structure of the Virgo Cluster.

\section*{Acknowledgments} 

This work was based on observations made through the Isaac Newton
Groups' Wide Field Camera Survey Programme with the Isaac Newton
Telescope operated on the island of La Palma by the Isaac Newton Group
in the Spanish Observatorio del Roque de los Muchachos of the
Instituto de Astrofisica de Canarias.
The contributions of J.~Davies and R.~McMahon in initiating the WFS
and the Virgo
part of the survey are acknowledged in particular. 
We are also grateful to the referee, Dr.~S.~Phillipps, for detailed
comments on the manuscript.
This research has made use of the NASA/IPAC Extragalactic Database (NED)
which is operated by the Jet Propulsion Laboratory, Caltech, under agreement
with the National Aeronautics and Space Association.

{}

\end{document}